\documentclass[lettersize,journal]{IEEEtran}

\usepackage{subfigure}
\usepackage{colortbl}
\usepackage{bm}

\usepackage{graphicx}  
\usepackage{url}       

\usepackage{amsmath}    
\allowdisplaybreaks[4]
\usepackage{cite}

\usepackage{amsfonts,amssymb}

\usepackage{stfloats}

\usepackage{cases}
\usepackage{algorithm}
\usepackage{multirow}
\usepackage{algorithmic}
\usepackage{epstopdf}
\usepackage{color}

\usepackage[square, comma, sort&compress, numbers]{natbib}

\makeatletter
\newcommand{\vvast}{\bBigg@{3.0}}
\newcommand{\vast}{\bBigg@{4}}
\newcommand{\Vast}{\bBigg@{4.5}}
\newcommand{\VVast}{\bBigg@{5}}
\newcommand{\VVVast}{\bBigg@{5.5}}
\newcommand{\VVVVast}{\bBigg@{6}}

\makeatother

\newcommand{\ls}[1]
{\dimen0=\fontdimen6\the\font
	\lineskip=#1\dimen0
	\advance\lineskip.5\fontdimen5\the\font
	\advance\lineskip-\dimen0
	\lineskiplimit=.9\lineskip
	\baselineskip=\lineskip
	\advance\baselineskip\dimen0
	\normallineskip\lineskip
	\normallineskiplimit\lineskiplimit
	\normalbaselineskip\baselineskip
	\ignorespaces
}

\ifCLASSINFOpdf
\else
\fi
\begin{document}

		\title{\ls{1.0}{AoI-Aware Resource Allocation for Smart Multi-QoS Provisioning}}

\author{\IEEEauthorblockN{Jingqing Wang,~\IEEEmembership{Member,~IEEE}, Wenchi Cheng,~\IEEEmembership{Senior Member,~IEEE}, and Wei Zhang,~\IEEEmembership{Fellow,~IEEE}} \\[0.2cm]
	
	\thanks{
		This work was supported in part by National Key R\&D Program of China under Grant 2021YFC3002100.}
	\thanks{Jingqing Wang and Wenchi Cheng are with the State Key Laboratory of Integrated Services Networks, Xidian University, Xi’an, China (e-mails: jqwangxd@xidian.edu.cn; wccheng@xidian.edu.cn).}
	\thanks{Wei Zhang is with the School of Electrical Engineering and Telecommunications, The University of New South Wales, Sydney, Australia (e-mail: w.zhang@unsw.edu.au).}
}

\maketitle


\begin{abstract}
The Age of Information (AoI) has recently gained recognition as a critical quality-of-service (QoS) metric for quantifying the freshness of status updates, playing a crucial role in supporting massive ultra-reliable and low-latency communications (mURLLC) services.
In mURLLC scenarios, status updates generally involve the transmission through applying finite blocklength coding (FBC) to efficiently encode these small update packets while meeting stringent error-rate and latency-bounded QoS constraints.
However, due to the inherent system dynamics and varying environmental conditions, optimizing AoI under such multi-QoS constraints considering both delay and reliability often results in non-convex and computationally intractable problems. Motivated by the demonstrated efficacy of deep reinforcement learning (DRL) in addressing large-scale networking challenges, this work aims to apply DRL techniques to derive optimal resource allocation solutions in real time. 
Despite its potential, the effective integration of FBC in DRL-based AoI optimization remains underexplored, especially in addressing the challenge of simultaneously upper-bounding both delay and error-rate.
To address these challenges, we propose a DRL-based framework for AoI-aware optimal resource allocation in mURLLC-driven multi-QoS schemes, leveraging AoI as a core metric within the finite blocklength regime. 
First, we design a wireless communication architecture and AoI-based modeling framework that incorporates FBC. 
Second, we proceed by deriving upper-bounded peak AoI and delay violation probabilities using stochastic network calculus (SNC). 
Subsequently, we formulate an optimization problem aimed at minimizing the peak AoI violation probability through FBC.
Third, we develop DRL algorithms to determine optimal resource allocation policies that meet statistical delay and error-rate requirements for mURLLC. 
Finally, to validate the effectiveness of the developed schemes, we have executed a series of simulations.
\end{abstract}


	\section{Introduction}\label{sec:intro}
	
	\IEEEPARstart{R}{esearchers} have developed the statistical delay-bounded quality-of-service (QoS)~\cite{10494937} for modeling queueing behaviors in response to the rapidly increasing demand for time-sensitive multimedia applications in emerging 6G mobile networks, as described in~\cite{10054381}. 
The dramatic increase in delay-sensitive multimedia traffic, coupled with stringent multi-QoS requirements, has intensified the need for stringent performance guarantees, such as bounded end-to-end delays (below 1 ms), ultra-high reliability (exceeding $>99.99999\%$), etc. These requirements pose considerable challenges for next-generation wireless communication systems, necessitating advanced approaches to ensure robust and efficient service delivery under stringent multi-QoS constraints.
	
One of the 6G usage scenarios, massive Ultra-Reliable Low-Latency Communications (mURLLC)~\cite{10177877,9635675}, has been developed to design and assess multi-QoS performances. To meet these requirements, researchers have investigated finite blocklength coding (FBC)~\cite{7529226,yury2010,Yp2014}, which reduces access delay while meeting mURLLC-driven multi-QoS needs~\cite{chen2020massive}.
The study in~\cite{yury2010} demonstrates that reliable communications can be achieved at the maximal channel coding rate for a given blocklength and non-vanishing error probability. Further, the work in~\cite{Yp2014} examines the properties of channel codes that approach the fundamental limits of a given memoryless wireless channel using FBC, contributing to the development of efficient communication techniques under stringent mURLLC constraints.
	

To ensure the timely and reliable transmission of age-sensitive data in fusion and communication systems, the concept of Age of Information (AoI)~\cite{10334482} has recently emerged as a novel QoS metric, specifically designed to quantitatively assess data freshness through monitoring the status updates. 
Given that status-update packets typically contain small data packets and must be delivered with minimal delay, 
FBC-based AoI measurement has been introduced as a means of maintaining data freshness, which plays a pivotal role in minimizing AoI by enabling rapid transmission and decoding of status update packets. 
Recent studies have identified a fundamental tradeoff between information staleness, represented by AoI, and transmission delays, including scheduling and queuing delays within the network. 
Unlike traditional transmission latency, information latency encompasses not only end-to-end delays but also factors such as communication reliability, sensing delays, information processing times, and other stages of the information acquisition process that collectively contribute to overall information latency~\cite{9023932}.	

Rather than focusing on average AoI metrics, increasing research attention has been directed toward analyzing the tail behavior of AoI, specifically the probability that AoI exceeds a predetermined threshold. This approach is particularly important for supporting mURLLC traffic, where stringent performance guarantees are required. The work in~\cite{8691802} investigates the probability of AoI outage for D/G/1 queuing systems, examining scenarios where the peak AoI exceeds a given threshold. Similarly, AoI outage probability for orthogonal multiple access (OMA) systems in the finite blocklength regime is studied in~\cite{9145084}.
In vehicular networks, a novel framework for characterizing and optimizing the tail distribution of AoI is proposed in~\cite{abdel2018ultra}. Additionally, the importance of focusing on AoI violation probabilities, rather than average AoI, is highlighted in~\cite{8437671}, which investigates both delay violation and peak-age violation probabilities. These studies highlight the critical role of AoI violation probabilities in ensuring ultra-reliable and low-latency communications.

There are two major challenges in addressing the minimization of AoI violation probabilities while ensuring stringent QoS requirements to support mURLLC traffic in the finite blocklength regime.
First, the AoI-driven multi-QoS metrics often do not have closed-form expressions. As a result, solving these optimization problems necessitates extensive simulations, which are computationally intensive and lead to significant processing delays~\cite{8402240}.
Second, AoI-driven optimization problems are inherently non-linear and non-convex, especially when considering the numerous dynamic aspects of wireless communication systems. 
The complexity, high dimensionality, and constantly evolving nature of these system states, action spaces, and environmental factors make many of these features difficult to model accurately in the existing literature.
Given these dynamic scenarios, the analysis and optimization of AoI-driven problems present significantly greater challenges compared to traditional communication delay optimization, particularly in massive access scenarios for mURLLC. These complexities highlight the need for advanced techniques capable of addressing the intricate and time-varying nature of next-generation wireless communication systems.

To address the challenges outlined above, deep reinforcement learning (DRL) has emerged as a promising technique due to its strong capability in solving large-scale networking problems. 
DRL is well-suited for solving complex Markov decision process (MDP) problems, leveraging deep neural networks (DNNs) as powerful non-linear approximation functions~\cite{sutton2018reinforcement}.
DRL-based algorithms have been extensively applied to dynamic optimization in wireless communication systems~\cite{8664581}, where the DRL agent continuously interacts with the environment to iteratively refine its policy based on reward feedback from the system~\cite{8743390}. 
A deep Q-network (DQN)-based framework was proposed to demonstrate significant energy cost reductions with only a slight compromise in average AoI performance.
In~\cite{10145068}, the authors explored a dynamic and learning-based scheduling approach to minimize the AoI in resource-constrained multi-source relaying systems. 
Despite these advances, previous studies on DRL-based mURLLC optimization problems have predominantly assumed infinite blocklength, without considering the stringent upper bounds on latency/error-rates. 
Applications of DRL to effectively implement FBC in the formulation and solution of AoI-driven optimization problems remain an open research question.

	To address the aforementioned problems, we apply the DRL algorithms for optimizing AoI-driven optimal resource allocations while guaranteeing multi-QoS provisioning through FBC.
	In particular, we develop FBC based system models, including the wireless communication model, channel coding rate model, AoI-metric based modeling frameworks, and random access protocols to support mURLLC. 
	Also, we characterize the upper-bounded AoI violation probability and delay violation probability functions through stochastic network calculus (SNC) to support the multi-QoS provisioning using FBC.
	Furthermore, we formulate the peak AoI violation probability minimization problem in the finite blocklength regime,  model the proposed minimization problem by MDP, and develop two novel DRL techniques, including double DQN (DDQN) and deep dueling neural network (DDNN) algorithms, to solve the AoI-aware optimization problem while supporting multi-QoS for mURLLC. 
	Finally, we validate and assess the system performance through a series of simulations.

	The rest of this paper is organized as follows: Section~\ref{sec:sys} builds AoI-metric based system model.
	Section~\ref{sec:EC} derives the upper-bounds for peak AoI and delay violation probabilities and formulates the peak AoI violation probability minimization problem under mURLLC constraints. 
	Section~\ref{sec:RL} applies DRL algorithms to search for the optimal resource allocation policy for mURLLC. 
	Section~\ref{sec:results} validates and assesses the system performance for the developed algorithms. This paper concludes with Section~\ref{sec:conclusion}.

	\section{The System Models}\label{sec:sys}
\begin{figure*}[!t]
	\centering
	\includegraphics[scale=0.52]{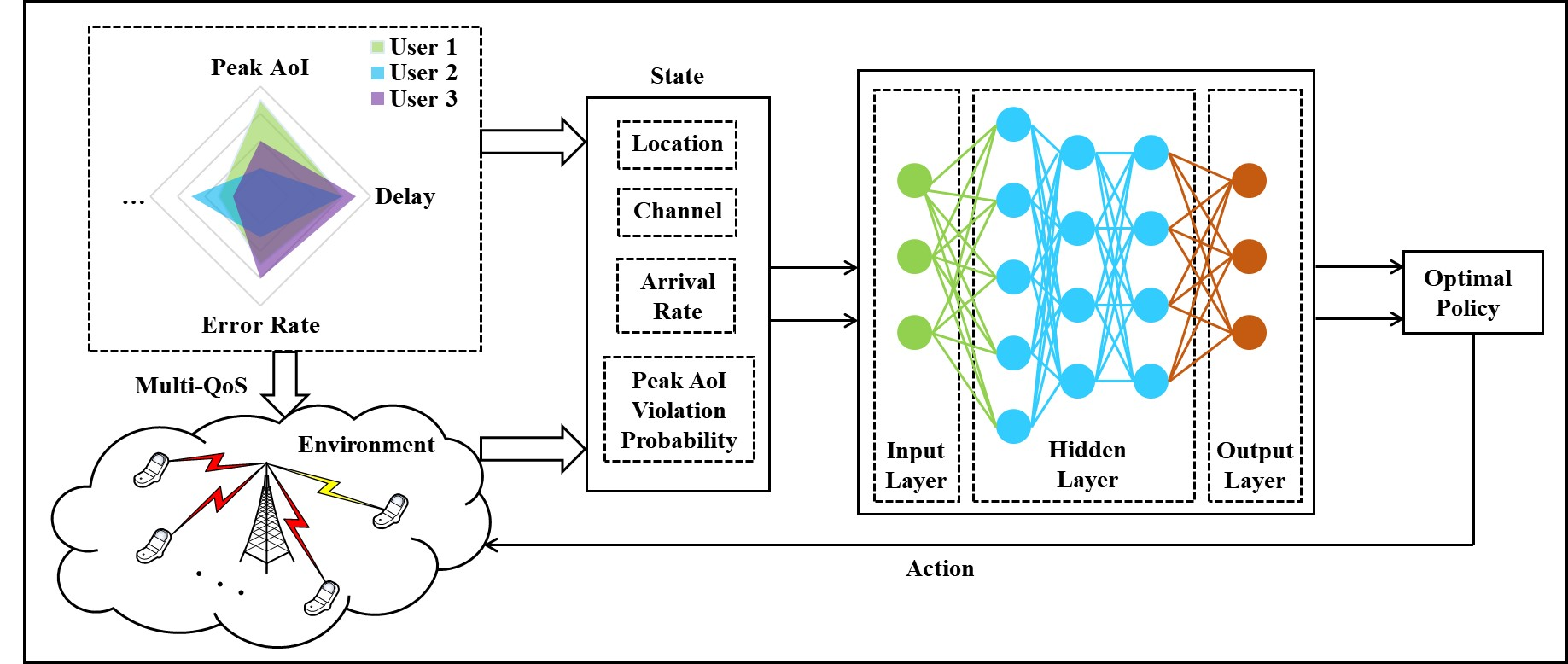}
	\caption{The learning-based AoI-aware resource allocation model for guaranteeing multi-QoS provisioning.}
	\label{fig:1}
\end{figure*}

Consider the network architecture model for the developed learning-based AoI-aware resource allocation schemes for guaranteeing multi-QoS in terms of delay and error-rate, as shown in Fig.~\ref{fig:1}.
Consider a scenario in which a single base station (BS) serves 
$K$ mobile devices (MDs), such as Internet of Things (IoT) sensors, tasked with transmitting telemetry (TM) status-update data packets to the BS for monitoring the AoI in time-sensitive applications, including vehicle-to-everything (V2X) and IoT environments. 
Since TM status update packets typically carry small amounts of data but must be transmitted with stringent low-latency and high-reliability requirements, the use of FBC is particularly suitable for ensuring efficient transmission while maintaining reliability guarantees essential for mURLLC.
	Denote by $\bm{q}_{k}^{(\mu)}=\left[x_{k}^{(\mu)}, y_{k}^{(\mu)}\right]$ the coordinate for the position of MD $k$ when transmitting the $\mu$th TM status update $(\mu=1,\dots, N)$, where $N$ is the total number of status-update packets.
	
	\subsection{The Wireless Channel Model}
	We consider block fading channel, where the channel remains constant within a coherence interval and varies independently among intervals. 
We assume a time-slotted system where radio resources are allocated in each time frame, referred to as the transmit time interval (TTI)~\cite{6226795}.
Denote by $T_{f}$ the duration of a time frame. 
While transmitting TM status update $\mu$, we define the binary variable $b_{k,l}(\mu)$ for \textit{subchannel allocation} for MD $k$ over the $l$th subchannel as follows:
	\begin{equation}
		b_{k,l}(\mu)=
		\begin{cases}
			1, \,\,\, \text{if MD $k$ is connected through subchannel $l$;}\\
			0, \,\, \,\text{otherwise},
		\end{cases}
	\end{equation}
while satisfying the following constraints: 
	\begin{align}
		\sum_{l=1}^{L}b_{k,l}(\mu)\leq1, \quad \forall k
		\quad \text{   and  } \quad
		\sum_{k=1}^{K}b_{k,l}(\mu)= K^{(\mu)}_{l}, \quad \forall l
	\end{align}
	where $L$ is the total available subchannels and $K^{(\mu)}_{l}$ is the total number of the MDs that share $l$th subchannel for transmitting data packet $\mu$.
The above constraint ensures that the resource allocation framework efficiently manages subchannel assignments, preventing any MD from being associated with more than one subchannel simultaneously, while allowing for flexible and dynamic sharing of subchannels across multiple MDs to optimize network performance.

	The overall communication process over the $l$th subchannel can be divided into two phases, i.e., control phase and transmission phase. 
	During the control phase, all MDs send request connections to the BS. The BS selects a joint access control and resource allocation policy, and then broadcasts to all MDs. This policy includes subchannel allocations and the corresponding access control measures.
	During the transmission phase, we assume that each $T_{f}$ ms-long frame is equally divided into $RB_{l}$ \textit{resource blocks} (RBs) over subchannel $l$.
	Denote by $B_{l}$ the bandwidth of each subchannel.
	A message-encoder at MD $k$ maps the message into a codeword with a length of $n_{l}$, where $n_{l}$ is defined as the blocklength of the RB over the $l$th subchannel. 
	For TM update packet $\mu$ $(\mu=1,\dots, N)$, MD $k$ encodes the data packet into a codeword with $n_{l}$ blocklength. 
	As a result, the number of resource blocks over the $l$th subchannel can be derived as follows:
	\begin{equation}
		RB_{l}=\frac{T_{f}B_{l}}{n_{l}}.
	\end{equation}
To optimize resource utilization, each MD is permitted to occupy a maximum of one RB for transmitting an update data packet within a single frame.
	
	We can derive the signal-to-interference-plus-noise ratio (SINR), denoted by $\gamma^{(\mu)}_{k,l}$, for transmitting the $\mu$th TM update data packet over the $l$th subchannel as follows:
	\begin{align}\label{equation010}
		\gamma^{(\mu)}_{k,l}=&\frac{{\cal P}_{k,l}^{(\mu)}\left(d_{k}^{(\mu)}\right)^{-\varsigma}\bm{h}_{k,l}^{(\mu)}\left(\bm{h}_{k,l}^{(\mu)}\right)^{H}}{\sum\limits_{j=1,j\neq k}^{K}{\cal P}_{j,l}^{(\mu)}\left(d_{j}^{(\mu)}\right)^{-\varsigma}\bm{h}_{j,l}^{(\mu)}\left(\bm{h}_{j,l}^{(\mu)}\right)^{H}+\sigma^{2}}
	\end{align}
	where $(\cdot)^{H}$ is the conjugate-transpose of a matrix, ${\cal P}_{k,l}^{(\mu)}$ and ${\cal P}_{j,l}^{(\mu)}$ are the transmit powers for the $k$th and $j$th MD, respectively, while transmitting TM status update $\mu$, $\varsigma$ is the path envelope, $\sigma^{2}$ is the noise power, $\bm{h}_{k,l}^{(\mu)}$ and $\bm{h}_{j,l}^{(\mu)}$ are the channel's impulse response vectors from the $k$th and $j$th MD, respectively, to the BS over the $l$th subchannel, which follow the Rayleigh fading model, and $d_{k}^{(\mu)}=\sqrt{\left\|\bm{q}_{k}^{(\mu)}\right\|^{2}}$ and $d_{j}^{(\mu)}=\sqrt{\left\|\bm{q}_{j}^{(\mu)}\right\|^{2}}$ are the distances between the BS and the $k$th and $j$th MD, respectively, where $\|\cdot\|$ is the Euclidean distance.

	\subsection{The Channel Coding Rate in the Finite Blocklength Regime}

	Traditional coding techniques assume infinitely long blocklengths, achieving asymptotically optimal performance. However, in practical scenarios--where TM packets must be transmitted with ultra-low latency--only a finite number of symbols can be used for encoding. 
	This necessitates a careful balance between transmission rate, latency, and decoding error probability. FBC provides a framework for optimizing this trade-off, enabling shorter blocklengths that reduce transmission time while adhering to the stringent multi-QoS demands of mURLLC.
	In particular, the $(n, M, \epsilon)$ code framework~\cite{yury2010} is used in FBC to represent the performance of coding schemes where $n$ represents the blocklength, which is the number of channel uses, $M$ represents the cardinality of message space, and $\epsilon$ is the decoding error probability.

	\textit{Definition 1. The Non-Vanishing Decoding Error Probability:}
The decoding error probability function, denoted by $\epsilon\left(\gamma_{k,l}^{(\mu)}\right)$, for transmitting TM status update $\mu$ over $l$th subchannel is derived as follows~\cite{7529226}:
	\begin{equation}\label{equation018}
		\epsilon\left(\gamma_{k,l}^{(\mu)}\right)\approx  {\cal Q}\left(\frac{C\left(\gamma_{k,l}^{(\mu)}\right)-\frac{\log_{2}(M)}{n_{l}}}{\sqrt{V\left(\gamma_{k,l}^{(\mu)}\right)/n_{l}}}\right)
	\end{equation}
	where ${\cal Q}(\cdot)$ denotes the \textit{$Q$}-function and $C\left(\gamma_{k,l}^{(\mu)}\right)$ and $V\left(\gamma_{k,l}^{(\mu)}\right)$ are the \textit{channel capacity}  and \textit{channel dispersion}, respectively, which are defined as follows:
	\begin{equation}
		\begin{cases}
			C\left(\gamma_{k,l}^{(\mu)}\right)=\log_{2}\left(1+\gamma_{k,l}^{(\mu)}\right); \\
			V\left(\gamma_{k,l}^{(\mu)}\right)=1-\frac{1}{\left(1+\gamma_{k,l}^{(\mu)}\right)^{2}}.
		\end{cases}
	\end{equation}
	
	\subsection{Random Access Protocol}
	
	When the number of RBs is sufficiently large, the BS can assign a dedicated RB to each MD, thereby avoiding packet collisions.
	However, considering mURLLC traffics, the scenario often involves a substantial number of MDs, with the total exceeding the available RBs. 
	In this scenario, due to the bursty nature of communications among MDs, random access protocols~\cite{8006544,8445979,9181539} have been proposed to facilitate synchronization between MDs and BSs.
MDs attempt to establish connections with the BS in an uncoordinated manner by randomly selecting preambles. This lack of coordination can lead to collisions when multiple devices choose the same preamble within the same time frame, resulting in significant delays and packet losses. To mitigate this issue, we implement the \textit{access control barring} (ACB) scheme~\cite{7404058,6398917}.
In the control phase, the BS broadcasts an access probability threshold, denoted as $p_{l,\text{access}}$, to all MDs over the $l$th subchannel. 
At the beginning of each time frame, the MD randomly generates a value $q\in(0,1)$ and attempts to access the $l$th subchannel only when $q\leq p_{l,\text{access}}$.
If the generated value exceeds this threshold, the MDs refrain from accessing the channel and will re-evaluate their access attempt in the subsequent time frame.
This approach ensures that if a single MD selects a specific RB, a collision is avoided. Conversely, if two or more MDs choose the same RB, a collision occurs, preventing any of the update data packets from being successfully transmitted to the BS.
Correspondingly, we derive the successfully access probability, denoted by $p_{l,\text{succ}}$, for each MD when it joins in the competition with other $(K^{(\mu)}_{l}-1)$ contenders over the $l$th subchannel as follows:
	\begin{align}\label{equation09}
		p_{l,\text{succ}}&=\sum_{k=1}^{K^{(\mu)}_{l}}\left(1-\frac{1}{RB_{l}}\right)^{k}\binom{K^{(\mu)}_{l}-1}{k}
		\left(p_{l,\text{access}}\right)^{k}
		\nonumber\\
		&\qquad\times \left(1-p_{l,\text{access}}\right)^{K^{(\mu)}_{l}-1-k}
		\nonumber\\
		&=\left(1-\frac{p_{l,\text{access}}}{RB_{l}}\right)^{K^{(\mu)}_{l}-1}
	\end{align}
	where $\binom{K^{(\mu)}_{l}-1}{k} = \frac{(K^{(\mu)}_{l}-1)!}{k!(K^{(\mu)}_{l}-1-k)!}$ is the Binomial coefficient.
	

	\section{The Modeling and Optimization for Peak AoI Violation Probability Using FBC}\label{sec:EC}
	
	
	\subsection{Peak AoI Violation Portability Analyses for mURLLC}
To effectively assess and manage data freshness, we utilize the peak AoI as a critical performance metric~\cite{10609803,10494937}. 
For a given designated peak AoI threshold, denoted by $A_{\text{th}}$, our objective is to characterize the tail behavior of the peak AoI by deriving the peak AoI violation probability, which is a crucial factor in determining the fundamental performance-limits of statistical QoS.
The peak AoI violation probability, denoted by {$p^{(\mu,\text{AoI})}_{k,l}$}, is defined as the probability that the peak AoI, denoted by $P_{k,l}^{\text{AoI}}(\mu)$, exceeds a threshold $A_{\text{th}}$, which is given as follows:
\begin{align}\label{equation49}
	{p^{(\mu,\text{AoI})}_{k,l}\triangleq\text{Pr}\left\{ P_{k,l}^{\text{AoI}}(\mu)>A_{\text{th}}\right\}.}
\end{align}
However, it is challenging to directly obtain the closed-form expression of peak AoI violation probability. Instead, we apply the Mellin transform to derive an effective upper bound. This approach enables us to analyze and constrain the peak AoI violation probability by transforming the problem into a more tractable form, facilitating the evaluation of system performance under stringent multi-QoS requirements.
Assume that the inter-arrival time and service time are independent and identically distributed (i.i.d.).
As shown in our previous work, the following upper-bounded peak AoI violation probability is expressed through applying the Mellin transform~\cite{JSAC_jingqing2021,9400231}:
	\begin{equation}\label{equation17}
		p^{(\mu,\text{AoI})}_{k,l}\leq e^{-\theta A_{\text{th}}}\frac{{\cal M}_{{\cal T}^{\text{I}}_{k,l}(\mu-1,\mu)}(1+\theta){\cal M}_{{\cal T}^{\text{S}}_{k,l}(\mu)}(1+\theta)}{1-{\cal M}_{{\cal T}^{\text{I}}_{k,l}(\mu-1,\mu)}(1-\theta){\cal M}_{{\cal T}^{\text{S}}_{k,l}(\mu)}(1+\theta)
		},
	\end{equation}
while the stability condition ${\cal M}_{{\cal T}^{\text{I}}_{k,l}(\mu-1,\mu)}(1-\theta){\cal M}_{{\cal T}^{\text{S}}_{k,l}(\mu)}(1+\theta)<1$ holds, where $\theta>0$ is the peak-AoI bounded QoS exponent and ${\cal M}_{{\cal T}_{k,l}^{\text{I}}(\mu-1,\mu)}(\theta)$ and ${\cal M}_{{\cal T}^{\text{S}}_{k,l}(\mu)}(\theta)$ are the Mellin transforms with respect to the inter-arrival time between TM update packets $(\mu-1)$ and $\mu$ and the service time for transmitting TM  status-update $\mu$, respectively, over the $l$th subchannel in the exponential domain.

{\textit{Definition 2:} Through applying the Large Deviation Principle (LDP), the asymptotic upper-bounded peak AoI violation probability is achieved as the peak AoI threshold becomes sufficiently large, i.e., as $A_{\text{th}}\rightarrow\infty$, we have
	\begin{align}\label{equation18}
	&	-\!\!\lim_{A_{\text{th}}\rightarrow\infty}\!\!\frac{\log\!\left(\text{Pr}\!
			\left\{ P_{k,l}^{\text{AoI}}(\mu)\!>\!A_{\text{th}}\right\}\!\right)}{A_{\text{th}}}
		=\theta
	\end{align}
	where $\theta$ measures the exponential decay rate of the peak AoI violation probability.
	The above equation implies that the probability of the peak AoI exceeding a certain threshold $A_{\text{th}}$ decays exponentially fast as the peak AoI threshold $A_{\text{th}}$ increases.

	Furthermore, given that the arrival process is modeled as a Poisson process with rate rate $\lambda_{k,l}$, the inter-arrival time follows an exponential process with rate $1/\lambda_{k,l}$. 
Accordingly, the Mellin transform of the inter-arrival time is expressed as follows:
	\begin{equation}\label{equation33}
		{\cal M}_{{\cal T}_{k,l}^{\text{I}}(\mu-1,\mu)}(1+\theta)=\frac{1}{1-\lambda_{k,l}\theta}.
	\end{equation}
Moreover, the Mellin transform of the service time when transmitting TM status-update $\mu$ is given as follows:
	\begin{equation}\label{equation196}
		{\cal M}_{{\cal T}^{\text{S}}_{k,l}(\mu)}(1+\theta)=\mathbb{E}\left[e^{\theta T^{\text{S}}_{k,l}(\mu)}\right]
		=\frac{\left[1-p_{l,\text{o}}(\mu)\right]e^{\theta T_{f}}}{1-p_{l,\text{o}}(\mu)e^{\theta T_{f}}} 
	\end{equation}
	where $p_{l,\text{o}}(\mu)$ is the overall successfully transmission probability for transmitting TM update packet $\mu$ from MD $k$ to the BS over subchannel $l$.
A status-update is successfully transmitted if it is not blocked by the access control protocol, does not experience packet collisions, and is successfully decoded without errors. Therefore, the overall transmission success probability is expressed as follows:	
	\begin{equation}\label{equation196b}
		p_{l,\text{o}}(\mu)= p_{l,\text{succ}}p_{l,\text{access}}\epsilon\left(\gamma_{k,l}^{(\mu)}\right)
	\end{equation}
	where $p_{l,\text{succ}}$ is given by Eq.~\eqref{equation09}.
	Note that Eqs.~\eqref{equation196} and~\eqref{equation196b} imply that the Mellin transform of the service time is an increasing function of the decoding error probability.
By plugging Eqs.~(\ref{equation33}),~\eqref{equation196},  and~\eqref{equation196b} into Eq.~(\ref{equation17}), we obtain
\begin{align}\label{equation35}
	p^{(\mu,\text{AoI})}_{k,l}&\leq \frac{\frac{\left[1-p_{l,\text{o}}(\mu)\right]e^{\theta T_{f}}e^{-\theta A_{\text{th}}}}{\left[1-\lambda_{k,l}\theta\right]\left[1-p_{l,\text{o}}(\mu)e^{\theta T_{f}}\right]} }{1-\frac{\left[1-p_{l,\text{o}}(\mu)\right]e^{\theta T_{f}}}{\left[1+\lambda_{k,l}\theta\right]\left[1-p_{l,\text{o}}(\mu)e^{\theta T_{f}}\right]} 
	}.
\end{align}

	\subsection{Delay Violation Portability Analyses for mURLLC}
	Statistical delay-bounded QoS theory~\cite{10494937} is investigated in the context of analyzing queueing behaviors, which is defined as follows.	
	
	{\textit{Definition 3:} Through applying the LDP, under sufficient conditions, the queueing process converges in distribution to a random variable $Q(\infty)$ such that
		\begin{equation}\label{equation23}
			-\lim_{Q_{\text{th}}\rightarrow\infty}\frac{\log\left(\text{Pr}
				\left\{Q(\infty)>Q_{\text{th}}\right\}\right)}{Q_{\text{th}}}=\widetilde{\theta}
		\end{equation}
where $Q_{\text{th}}$ is the overflow threshold and $\widetilde{\theta}>0$ is defined as the delay-bounded \textit{QoS exponent}, measuring the exponential decay rate of the delay violation probability.
Similar to the peak-AoI bounded QoS exponent, the above equation implies that the probability of the delay violation exceeding a certain threshold $Q_{\text{th}}$ decays exponentially fast as $Q_{\text{th}}$ increases. 

For our proposed smart multi-QoS schemes, the concept of the  \textit{$\epsilon$-effective capacity} is proposed to provide multi-QoS guarantees in terms of delay and error-rate through FBC.
Accordingly, the service process, denoted by $S_{k,l}(\mu)$, when delivering TM status update $\mu$ over subchannel $l$ is derived as follows:			
	{ \begin{equation}\label{equation021}
			S_{k,l}(\mu)=
			\begin{cases}
				\log_{2}(M), \,\,\quad\text{with probability } 1-\epsilon\left(\gamma_{k,l}^{(\mu)}\right);\\
				0,  \qquad\qquad\,\,\, \text{with probability } \epsilon\left(\gamma_{k,l}^{(\mu)}\right).
			\end{cases}
		\end{equation}
	In addition, the Mellin transform of the service process, denoted by ${\cal M}_{{\cal S}_{k,l}(\mu)}\left(\widetilde{\theta}\right)$, in the exponential domain is expressed as follows:
	\begin{align}\label{equation31}
		&{\cal M}_{{\cal S}_{k,l}(\mu)}\left(\widetilde{\theta}\right)\!
		=\mathbb{E}_{\gamma_{k,l}^{(\mu)}}\left[e^{\left(\widetilde{\theta}-1\right)  S_{k,l}(\mu)}\right] 
		\nonumber\\
		&\quad=\mathbb{E}_{\gamma_{k,l}^{(\mu)}}\left[\epsilon\left(\gamma_{k,l}^{(\mu)}\right)\right] +\mathbb{E}_{\gamma_{k,l}^{(\mu)}}\left[1\!-\!\epsilon\left(\gamma_{k,l}^{(\mu)}\right)\right] \!
		e^{\left(\widetilde{\theta}-1\right)  \log_{2}(M)}
	\end{align}
where $\mathbb{E}_{\gamma_{k,l}^{(\mu)}}[\cdot]$ is the expectation operation over the SINR.
	%
Correspondingly, the expression of the $\epsilon$-effective capacity is defined as follows.
	
	\textit{Definition 4:} The \textit{$\epsilon$-effective capacity}, denoted by {$EC_{k,l}^{(\epsilon,\mu)}\left(\widetilde{\theta}\right)$}, is defined as the maximum constant arrival rate for a given service process, taking into account multi-QoS constraints in terms of delay and error-rate, which is formally expressed as:
	\begin{align}\label{equation017}
		EC_{k,l}^{(\epsilon,\mu)}\left(\widetilde{\theta}\right)&\triangleq-\frac{1}{n\widetilde{\theta}}\log\left\{{\cal M}_{{\cal S}_{k,l}(\mu)}\left(1-\widetilde{\theta}\right)\right\}.
	\end{align}
	

In practical scenarios, obtaining the statistical characteristics of random arrival and service processes poses significant challenges. 
To address this, considering the delay-sensitive services, the kernel function ${\cal K}\left(\widetilde{\theta}, D_{\text{th}}\right)$ is defined as follows~\cite{HAL2016}:
	\begin{equation}\label{equation26}
		{\cal K}\left(\widetilde{\theta}, D_{\text{th}}\right)\triangleq \frac{\left[{\cal M}_{{\cal S}_{k,l}(\mu)}\left(1-\widetilde{\theta}\right)\right]^{D_{\text{th}}}}{1-{\cal M}_{{\cal A}_{k,l}(\mu)}\left(1+\widetilde{\theta}\right){\cal M}_{{\cal S}_{k,l}(\mu)}\left(1-\widetilde{\theta}\right)},
	\end{equation}
	if the following stability condition holds:
	\begin{equation}
		{\cal M}_{{\cal A}_{k,l}(\mu)}\left(1+\widetilde{\theta}\right){\cal M}_{{\cal S}_{k,l}(\mu)}\left(1-\widetilde{\theta}\right)<1,
	\end{equation}
	where $D_{\text{th}}$ represents the delay bound and ${\cal M}_{{\cal A}_{k,l}(\mu)}\left(\widetilde{\theta}\right)$ represents the Mellin transform over the arrival process $A_{k,l}(\mu)$ in the exponential domain.
	Then, given the delay bound $D_{\text{th}}$, an upper-bounded delay violation probability, denoted by $p^{(\mu,\text{q})}_{k,l}$, is expressed through applying the kernel function ${\cal K}\left(\widetilde{\theta}, D_{\text{th}}\right)$ as follows:
	\begin{equation}\label{equation27}
		p^{(\mu,\text{q})}_{k,l}\leq  \inf_{\widetilde{\theta}>0}\left\{{\cal K}\left(\widetilde{\theta}, D_{\text{th}}\right)\right\}.
	\end{equation}
Assuming that arrivals during each time frame are i.i.d., the accumulated arrival rate $A_{k,l}(\mu)$ exhibits i.i.d. increments, represented as $\widetilde{a}_{k,l}(\mu)$ (or simply $\widetilde{a}_{k,l}=\widetilde{a}_{k,l}(\mu)$, given the i.i.d. nature of $\widetilde{a}_{k,l}(\mu)$. 	
Consequently, the Mellin transform over $A_{k,l}(\mu)$, denoted by ${\cal M}_{{\cal A}_{k,l}(\mu)}(\widetilde{\theta})$, is given as follows:
	\begin{align}\label{equation031}
		{\cal M}_{{\cal A}_{k,l}(\mu)}(\widetilde{\theta})&=\mathbb{E}\left[\left(\prod_{j=1}^{\mu}e^{\widetilde{a}_{k,l}(j)}\right)^{\widetilde{\theta}-1}\right]
		=\left(\mathbb{E}\left[e^{\widetilde{a}_{k,l}\left(\widetilde{\theta}-1\right)}\right]\right)^{\mu} \nonumber\\
		&=\left[{\cal M}_{\alpha_{k,l}}(\widetilde{\theta})\right]^{\mu}
	\end{align}	
	where $\alpha_{k,l}=e^{\widetilde{a}_{k,l}}$. 
Given that we assume the arrival process follows a Poisson distribution with an average rate of $\lambda_{k,l}$, the Mellin transform of $\alpha_{k,l}$ is given as follows:
	\begin{align}\label{equation24}
		{\cal M}_{\alpha_{k,l}}(\widetilde{\theta})=\sum_{i=1}^{\infty}e^{i(\widetilde{\theta}-1)}\frac{\left(\lambda_{k,l}\right)^{i}}{i!}e^{-\lambda_{k,l}}=e^{\lambda_{k,l}\left(e^{\widetilde{\theta}-1}-1\right)}.
	\end{align}	
	Therefore, based on Eqs.~\eqref{equation26},~\eqref{equation27}, and~\eqref{equation24}, we can rewrite the delay violation probability function as follows:
	\begin{align}\label{equation25}
		p^{(\mu,\text{q})}_{k,l}\leq  \inf_{\widetilde{\theta}>0}\left\{\frac{\left[{\cal M}_{{\cal S}_{k,l}(\mu)}\left(1-\widetilde{\theta}\right)\right]^{D_{\text{th}}}}
		{1-e^{\mu\lambda_{k,l}\left(e^{\widetilde{\theta}}-1\right)}
			{\cal M}_{{\cal S}_{k,l}(\mu)}\left(1-\widetilde{\theta}\right)}\right\}
	\end{align}
	which implies that the delay violation probability is monotonically increasing in the Mellin transform of the service process.
	We can demonstrate that the Mellin transform of the service process is monotonically increasing with respect to the decoding error probability.
	Consequently, both the Mellin transform of the service process and the Mellin transform of the service time are monotonically increasing in terms of the decoding error probability.
	Furthermore, by plugging Eq.~\eqref{equation017} into Eq.~\eqref{equation25}, we can obtain the following equation:
	\begin{align}\label{equation28}
		p^{(\mu,\text{q})}_{k,l}
	\!\leq \! \inf_{\widetilde{\theta}>0}\left\{\!\frac{\exp\left[-n\widetilde{\theta}D_{\text{th}}EC_{k,l}^{(\epsilon,\mu)}\left(\widetilde{\theta}\right)\right]}
		{1\!-\!		\exp\left[\lambda_{k,l}\left(e^{\widetilde{\theta}}-1\right)-n\widetilde{\theta}D_{\text{th}}EC_{k,l}^{(\epsilon,\mu)}\left(\widetilde{\theta}\right)\right]}\!\right\}\!.
	\end{align}
	The expression in Eq.~\eqref{equation28} constitutes the basis for the provisioning of statistical delay-bounded QoS guarantees in a wireless communication system in terms of the $\epsilon$-effective capacity function $EC_{k,l}^{(\epsilon,\mu)}\left(\widetilde{\theta}\right)$ and the QoS exponent of queuing delay $\widetilde{\theta}$.
	
	\subsection{Peak AoI Violation Probability Optimization for Multi-QoS Using FBC}
	
	Denote by $\bm{b}\triangleq \left\{b_{k,l}(\mu), \forall l\in\textsf{L},\forall \mu\in\textsf{N},\forall k\in\textsf{K}\right\}$, $\bm{{\cal P}}\triangleq \left\{{\cal P}^{(\mu)}_{k,l}, \forall l\in\textsf{L},\forall \mu\in\textsf{N}, \forall k\in\textsf{K}\right\}$, and 
	$\bm{p}_{\text{access}}\triangleq \left\{p_{l,\text{access}}, \forall l\in\textsf{L}\right\}$
	the subchannel allocation vector, transmit power vector, and access probability threshold vector, respectively, where 	$\textsf{N}\triangleq\{1,\dots, N\}$, $\textsf{K}\triangleq\{1,\dots, K\}$, and $\textsf{L}\triangleq\{1,\dots, L\}$ are the index sets for  $N$ TM states updates, $K$ MDs, and $L$ subchannels with their cardinalities: $|\textsf{N}|=N$, $|\textsf{K}|=K$, and $|\textsf{L}|=L$.
	To characterize tail behaviors of AoI metric--specifically, the peak AoI violation probability--while supporting mURLLC traffics, we formulate a minimization problem focused on the peak AoI violation probability through FBC.
	Given the peak AoI threshold $A_{\text{th}}$, the total peak AoI violation probability minimization problem $\mathbf{P_{1}}$ when delivering $N$ update data packets in our proposed schemes is formulated as follows:
	\begin{align}\label{equation036}
		&\mathbf{P_{1}:}
		\arg\min_{\left\{\bm{b},\bm{{\cal P}},\bm{p}_{\text{access}}\right\}} \left\{\sum_{l=1}^{L}\sum_{\mu=1}^{N}\sum_{k=1}^{K}p^{(\mu,\text{AoI})}_{k,l}\right\}
	\end{align}
	\begin{align}\label{equation190}
		&\text{s.t.} \quad \!\!\!C1\!: 	
		EC_{k,l}^{(\epsilon,\mu)}\left(\widetilde{\theta}\right)\geq EC_{\text{th}},
		 \quad k\in\textsf{K},l\in\textsf{L}; \\
		&\qquad\!\! C2\!: 		\mathbb{E}_{\gamma_{k,l}^{(\mu)}}\left[\sum_{k=1}^{K}{\cal P}_{k,l}^{(\mu)}\right]\leq\overline{{\cal P}}, \quad {\cal P}_{k,l}^{(\mu)}>0, \quad  k\in\textsf{K},l\in\textsf{L}; \label{equation192} \\
		&\qquad\!\! C3\!:	
		\sum_{l=1}^{L}b_{k,l}(\mu)\leq1, \quad k\in\textsf{K}; \label{equation193}\\
		&\qquad\!\! C4\!: 	\sum_{k=1}^{K}b_{k,l}(\mu)= K^{(\mu)}_{l}, \quad l\in\textsf{L}; \label{equation194}\\
		&\qquad\!\! C5\!: 0\leq p_{l,\text{access}}\leq 1, \quad l\in\textsf{L}, \label{equation195}
	\end{align}
	where $EC_{\text{th}}$ represents the upper bound on the $\epsilon$-effective capacity necessary to satisfy the queuing delay requirements  $\left\{D_{\text{th}},p^{(\mu,\text{q})}_{k,l}\right\}$ and $\overline{{\cal P}}$ represents the average transmit power constraint.
	Based on Eq.~\eqref{equation017}, the constraint $C1$ given by Eq.~\eqref{equation190} is converted as:
	\begin{equation}\label{equation197}
	C1': 	{\cal M}_{{\cal S}_{k,l}(\mu)}\left(1-\widetilde{\theta}\right)-e^{\widetilde{\theta}nEC_{\text{th}}}\leq 0, \quad k\in\textsf{K}.
	\end{equation}
	Since we observe that a a lower peak AoI violation probability correlates with increased transmit power, the optimal solution for $\mathbf{P_{1}}$ is attained when equality holds in constraint $C1'$. 
	Consequently, the $\mathbf{P_{1}}$ can be transformed into the equivalent Lagrangian problem $\mathbf{P_{2}}$ as follows:
	\begin{align}\label{equation037}
		\mathbf{P_{2}:}
		\arg\min_{\left\{\bm{b},\bm{{\cal P}},\bm{p}_{\text{access}}\right\}} &\Bigg\{\sum_{l=1}^{L}\sum_{\mu=1}^{N}\sum_{k=1}^{K}p^{(\mu,\text{AoI})}_{k,l}+
		\sum_{l=1}^{L}\sum_{\mu=1}^{N}\sum_{k=1}^{K}\eta_{k,l}^{(\mu)}
		\nonumber\\
		&\quad\!\times\! \left[{\cal M}_{{\cal S}_{k,l}(\mu)}\left(1\!-\!\widetilde{\theta}\right)\!-e^{\widetilde{\theta}nEC_{\text{th}}}\right]\!\!\Bigg\}
	\end{align}
	subject to constraints $C2-C5$ specified by Eqs.~(\ref{equation192})--(\ref{equation195}), respectively, where the first term on the right-hand side denotes the peak AoI violation probability and  
	$\eta_{k,l}^{(\mu)}$  is the Lagrangian multiplier.
	Since the subchannel allocation for each MD is a binary variable and the transmit power and access probability threshold $p_{l,\text{access}}$ are continuous variables, the minimization problem $\mathbf{P_{2}}$ in Eq.~(\ref{equation037}) is Mixed Integer Non-Linear Programming (MINLP), which makes it a non-convex problem.
	Integer-constrained problems even in the deterministic settings are known to be NP hard in general.
	Furthermore, all the MDs are competing over the available subchannels. 
	Due to the complexity and non-convexity of $\mathbf{P_{2}}$, especially considering the system dynamics and environment evolution over mURLLC-driven mobile wireless networks, it is intractable to solve the problem directly.
	Therefore, we propose DRL-based algorithms in order to learn and solve the proposed optimization problem $\mathbf{P_{2}}$ in the next section.

	\section{DRL-Based Resource Allocation Optimization for Multi-QoS Provisioning Through FBC}\label{sec:RL}
	
	In this section, we consider DRL-based algorithms to search for optimal resource allocation policies from real observations. 
	An MDP will be modeled to address the AoI-aware optimization problem. Then, we apply DRL approaches, including DDQN and DDNN algorithms, to solve the MDP problem in supporting statistical delay and error-rate bounded QoS provisioning in the finite blocklength regime for mURLLC. 

	\subsection{MDP Formulation for DRL}

	Since the above-mentioned optimization problem $\mathbf{P_{2}}$ is NP-hard, in order to select an optimal resource allocation strategy, we need to adopt the MDP framework to facilitate the selection of an optimal resource allocation strategy. 
	Typically, an MDP can be described by a tuple $({\textsf S},{\textsf A},{\textsf P},\textsf{R},\beta)$, where ${\textsf S}$ is the state space containing all possible network states, denoted by $\bm{s}(\mu)$, ${\textsf A}$ is the action space representing the collection of all possible actions, denoted  as $\bm{a}(\mu)$, ${\textsf P}$ is the transition probabilities $\text{Pr}\left\{\bm{s}(\mu+1)|\bm{s}(\mu),\bm{a}(\mu)\right\}$, ${\textsf R}$ is a reward received by the agent after conducting an action, and $\beta\in[0, 1)$ is the discount factor, which determines the
	importance of long-term reward. 
	For addressing the AoI-aware resource allocation optimization problem $\mathbf{P_{2}}$, we first define the following fundamental elements of the MDP:
	\begin{itemize}
		\item \textbf{State}: For update data packet $\mu$, we define the system state as follows:
		\begin{align}
			\bm{s}(\mu)&=\bigg(\left\{\bm{q}_{k}^{(\mu)}\right\}_{k\in\textsf{K},\mu\in\textsf{N}},
			\left\{\bm{h}_{k,l}^{(\mu)}\right\}_{k\in\textsf{K},\mu\in\textsf{N}},
			\left\{\lambda_{k,l}\right\}_{k\in\textsf{K},l\in\textsf{L}},
			\nonumber\\
			&\qquad\,\,
			\left\{p^{(\mu,\text{AoI})}_{k,l}\right\}_{k\in\textsf{K},l\in\textsf{L},\mu\in\textsf{N}}\bigg)\!\in\!{\textsf S}
			\nonumber\\
			&\triangleq {\cal Q}\times {\cal H}\times\Lambda\times {\mathcal Ao\mathcal I}
		\end{align}
		which includes the information of the geographical location $\bm{q}_{k}^{(\mu)}$, channel state $\bm{h}_{k,l}^{(\mu)}$, packet arrival rate $\lambda_{k,l}$, and peak AoI violation probability $p^{(\mu,\text{AoI})}_{k,l}$.
		We assume that the cardinalities of the packet arrival rate set, denoted by $|{\Lambda}|$, and the peak AoI violation probability set, denoted by $|\mathcal Ao\mathcal I|$, are finite, but can be arbitrarily large.

		\item \textbf{Action}: 
		The action is to determine the subchannel allocation variable $b_{k,l}(\mu)$ and  the access probability threshold $p_{l,\text{access}}$. 
		Define $\bm{a}(\mu)=
		\left(\left\{b_{k,l}(\mu)\right\}_{k\in\textsf{K},l\in\textsf{L}},
		\left\{p_{l,\text{access}}\right\}_{k\in\textsf{K}}\right)\in{\textsf A}$ as the actions for all $K$ MDs associated with $L$ subchannels for transmitting the $\mu$th data packet and  the access probability threshold.
		Define the set of available actions, denoted by $\textsf{A}$, performed for transmitting status update $\mu$ as follows:
		\begin{align}
			\textsf{A}:=\bigg\{&\bm{a}(\mu):\sum_{l=1}^{L}b_{k,l}(\mu)\leq1, \,\, \forall k
			\,\,\,\,
			\sum_{k=1}^{K}b_{k,l}(\mu)= K^{(\mu)}_{l}, 
			\nonumber\\
			&\qquad\qquad\quad \forall l,\,\,
			0\leq p_{l,\text{access}}\leq 1, \,\, l\in\textsf{L}\bigg\}.
		\end{align}

		\item \textbf{Reward Function}: The agent in an MDP receives a reward signal, denoted by $R(\mu+1)$, upon transitioning to the next state $\bm{s}(\mu+1)$. 
		Recall that we aim to minimize the peak AoI violation probability while guaranteeing mURLLC reliability constraint.
		Thus, we define the reward function, denoted by $R(\mu)$, as follows:
		\begin{align}\label{equation21a}
			R(\mu)=&-\sum_{l=1}^{L}\sum_{\mu=1}^{N}\sum_{k=1}^{K}p^{(\mu,\text{AoI})}_{k,l}-\sum_{l=1}^{L}\sum_{\mu=1}^{N}\sum_{k=1}^{K}\eta_{k,l}^{(\mu)}
			\nonumber\\
			& \times \left[{\cal M}_{{\cal S}_{k,l}(\mu)}\left(1-\widetilde{\theta}\right)-e^{\widetilde{\theta}nEC_{\text{th}}}\right].
		\end{align}
	\end{itemize}

The objective of the MDP-based approach is to identify the optimal resource allocation policy for each state, thereby maximizing the associated reward function.
For each update data packet, the agent observes some state $\bm{s}(\mu)$ and performs an action $\bm{a}(\mu)$. 
	After performing the action, the state of the environment transitions to $\bm{s}(\mu+1)$, and the agent receives a reward $R(\mu)$. 
	Let $\bm{\pi}=\left(\pi_{b},\pi_{{\cal P}},\pi_{p_{\text{access}}}\right)$ be a stationary control policy employed by the BS, where
	$\pi_{b}$, $\pi_{{\cal P}}$, and $\pi_{p_{\text{access}}}$ are the subchannel allocation policy, power allocation policy, and access probability threshold scheduling policy, respectively. 
	We aim to determine an optimal policy, denoted by $\bm{\pi}^{*}$, that maximizes the following expected discounted cumulative reward function:  
	\begin{equation}\label{equation303}
		\widetilde{R}\triangleq \sum_{\mu=1}^{\infty}\beta^{\mu} R(\mu).
	\end{equation} 
	We can apply the dynamic programming to derive the optimal resource allocation policy for the MDP based problem. 
	However, due to the large state space and stochastic nature of the wireless environment in practical scenarios, it is often infeasible to derive an exact model of the transition probability $\text{Pr}\left\{\bm{s}(\mu+1)|\bm{s}(\mu),\bm{a}(\mu)\right\}$ for our proposed mURLLC-driven multi-QoS schemes.
	Therefore, we apply the DRL technique which enables the MDs to explore various actions across dynamically changing network states over an extensive series of trials. This exploration allows the system to gradually adapt to fluctuating environments through real-time feedback.
	Through iterative learning and experience accumulation, the DRL framework enhances decision-making processes, enabling MDs to optimize their actions in response to fluctuating network conditions.

	\subsection{DRL-Based Resource Allocation Optimization}
	
Unlike the conventional RL methods  that consider only the most recent observation, denoted by $O^{(\mu-1)}$, 
the state variable $\bm{s}(\mu)$ for a DRL agent associated with the update packet $\mu$ comprises information from the preceding $T_{\text{o}}$ observations, i.e.,  $\bm{s}(\mu)=\left[O^{(\mu-T\text{o})},O^{(\mu-T\text{o}+1)},\dots, O^{(\mu-1)}\right]$.
This historical information is essential for capturing time-correlated features inherent in the traffic generation mechanisms and random access schemes. To identify patterns within the data packets, we employ a Recurrent Neural Network (RNN), specifically a Gated Recurrent Unit (GRU) network, to approximate the Q-function $Q(\bm{s}(\mu),\bm{a}(\mu))$ of each DRL algorithm.
Let $\bm{s}(0)\in\textsf{S}$ denote the initial state.
For each state $\bm{s}\in\textsf{S}$, we can obtain the expected value function, denoted by $\widetilde{V}(\bm{s})$, as follows:
	\begin{align}\label{equation302}
		\widetilde{V}(\bm{s})&=\mathbb{E}_{\bm{\pi}}\left[\sum_{\mu=0}^{\infty}\beta^{\mu} R(\mu)|\bm{s}(0)=\bm{s}\right]
		\nonumber\\
		&=\mathbb{E}_{\bm{\pi}}\left[ R(\mu)+\beta\widetilde{V}(\bm{s}(\mu+1))|\bm{s}(0)=\bm{s}\right].
	\end{align}
To determine the optimal policy $\bm{\pi}^{*}$, we identify the optimal action for each state by leveraging the optimal expected value function, denoted
by $\widetilde{V}^{*}(\bm{s})$, which is given as follows:
	\begin{equation}
		\widetilde{V}^{*}(\bm{s})=\arg\max_{\bm{a}\in\textsf{A}}\left\{\mathbb{E}_{\bm{\pi}}\left[ R(\mu)+\beta\widetilde{V}(\bm{s}(\mu+1))\right]\right\}, \quad \forall \bm{s}\in\textsf{S}.
	\end{equation}
	Correspondingly, we can derive the optimal Q-function, denoted by $Q^{*}(\bm{s},\bm{a}(\mu))$, for state-action pairs as follows:
	\begin{equation}
		Q^{*}(\bm{s},\bm{a}(\mu))=R(\mu)+\beta\mathbb{E}_{\bm{\pi}}\left[\widetilde{V}(\bm{s}(\mu+1))\right], \quad \forall \bm{s}\in\textsf{S}.
	\end{equation}
	The action selection is informed by the Q-function $Q(\bm{s}(\mu),\bm{a}(\mu))$ based on the observed $\bm{s}(\mu)$. 
	Utilizing the cumulative discounted reward function $\widetilde{R}(t)$, the DRL agent updates the Q-function  $Q(\bm{s}(\mu),\bm{a}(\mu))$ for action $\bm{a}(\mu)$ in an online manner, facilitating the identification of the optimal resource allocation policy for every state.
Historical observations in $O^{(\mu)}$ are sequentially input into the network, with the RNN maintaining connectivity to the output layer solely through the last input $O^{(\mu-1)}$ to inform action choices for data packet $\mu$.
It is important to note that the memory of the GRU RNN must be reinitialized at the beginning of each time frame. Additionally, the historical length $T_{\text{o}}$ should be appropriately selected to align with the expected memory requirements for recognizing time-correlated features.

The input consists of the variables in the state $\bm{s}(\mu)$. 
The intermediate layers are constructed using the GRU RNN, while the output layer is tailored to accommodate the specific requirements of various DRL algorithms. Let $\bm{\omega}$ represent the weight matrix of the GRU RNN.
In accordance with the distinct training principles of each DRL algorithm, a loss function, denoted by $\bm{{\cal L}}\left(\bm{\omega}^{(\mu)}\right)$, is computed to facilitate the update of the value function approximator $Q(\bm{s}(\mu),\bm{a}(\mu)|\bm{\omega})$.
Since the GRU RNN serves as the intermediate layers, the standard Stochastic Gradient Descent (SGD), implemented through BackPropagation Through Time (BPTT)~\cite{10097504}, is employed to perform the updates as follows:
	\begin{equation}
		\bm{\omega}^{(\mu+1)}=\bm{\omega}^{(\mu)}-\widehat{\alpha}\nabla\bm{{\cal L}}\left(\bm{\omega}^{(\mu)}\right)
	\end{equation}
	where $\widehat{\alpha}$ denotes the learning rate for SGD and $\nabla\bm{{\cal L}}\left(\bm{\omega}^{(\mu)}\right)$ denotes the gradient of the loss function for each DRL algorithms. 
	We consider the following two DRL algorithms, including DDQN and DDNN algorithm. 
	The learning principle of these two DRL algorithms are described as follows.
	

	\subsubsection{DDQN Algorithm}
The DDQN is proposed to enhance the performance of traditional deep Q-learning algorithm~\cite{9763399}. 
The fundamental concept behind DDQN is the sequential selection of actions. 
Initially, an action is chosen utilizing the primary network, which serves to determine the optimal action based on the current state. Subsequently, the target Q-value corresponding to the selected action is computed using the target network. This dual-network framework is instrumental in stabilizing the training process by mitigating the overestimation bias commonly encountered in traditional Q-learning algorithms.
Within this framework, the DDQN agent is designed to learn a state-action value function approximator, represented as $Q\left(\bm{s}(\mu),\bm{a}(\mu)|\bm{\omega}_{\text{DDQN}}^{(\mu)}\right)$, where $\bm{\omega}_{\text{DDQN}}$ represents the weight matrix of the DDQN. 
The weight matrix $\bm{\omega}_{\text{DDQN}}$ is updated in a fully online manner during each frame to circumvent the complexities associated with eligibility traces. Following the DDQN training principles, the gradient of the loss function $\bm{{\cal L}}\left(\bm{\omega}_{\text{DDQN}}^{(\mu)}\right)$, utilized for training the Q-function approximator,  is given as follows:
	\begin{align}\label{equation200}
		\nabla\bm{{\cal L}}\left(\bm{\omega}_{\text{DDQN}}^{(\mu)}\right)
		\!=&\,
		\mathbb{E}_{\bm{a}(\mu),\bm{s}(\mu),\bm{s}(\mu+1),R(\mu+1)}
		\bigg[\bigg(R(\mu+1)+\beta
		\nonumber\\
		&\times \! \max_{\bm{a}\in\textsf{A}}\left\{Q\left(\bm{s}(\mu+1),\bm{a}|\overline{\bm{\omega}}_{\text{DDQN}}^{(\mu)}\right)\right\}\bigg)\bigg]
	\end{align}
	where $\overline{\bm{\omega}}_{\text{DDQN}}^{(\mu)}$ is the target value function.
In the DDQN algorithm, the parameters of the target network  $\overline{\bm{\omega}}_{\text{DDQN}}^{(\mu)}$ are partially updated from the primary network parameters $\bm{\omega}_{\text{DDQN}}^{(\mu)}$ at each time frame.	
Let $\psi\in(0,1)$ represent the updating rate, while $I$ represents the total number of the iterations.
This updating rate is crucial for balancing the convergence speed of the learning process and the stability of the Q-value estimations, thereby facilitating the effective adaptation of the agent to the dynamic network environment throughout the iterative training process.

To mitigate the risk of overfitting during the training process, we explore a greedy policy, as suggested by~\cite{sutton2018reinforcement}. 
Specifically, during each iteration of the DDQN algorithm, the optimal action for a given state is selected with probability $(1-\tau)$, and a random action is sampled with probability $\tau$.
The expectation in Eq.~\eqref{equation200} is calculated over a randomly selected minibatch uniformly sampled from a finite replay memory, denoted as $\Gamma$, with size $|\Gamma|$. 
To reduce the correlation among observed sequences and enhance the overall stability, the experience replay technique~\cite{LI2024127204} is employed. In this method, system transition tuple $\left(\bm{a}(\mu),\bm{s}(\mu),\bm{s}(\mu+1),R(\mu+1)\right)$ is stored in ${\Gamma}$ following each training iteration. During subsequent iterations, a minibatch of transition tuples 
$\left(\bm{a}(j),\bm{s}(j),\bm{s}(j+1),R(j+1)\right)$ is randomly drawn from the replay memory ${\Gamma}$. Batch gradient descent is then employed to minimize the loss functions associated with this minibatch of transition tuples. This approach allows for more efficient utilization of previous experiences, enabling the algorithm to learn from them repeatedly.
Thus, the DDQN-based AoI-aware resource allocation algorithm is proposed, as outlined in
\textbf{Algorithm~\ref{alg1}}, to address $\mathbf{P_{2}}$. 
Note that although \textbf{Algorithm~\ref{alg1}} is presented in an online setting, the training process can be performed offline given that the proposed DDQN algorithm is based on an off-policy principle.

	\begin{algorithm} [!tp]
		\algsetup{linenosize=\tiny} \small
		\caption{DDQN-Based AoI-aware Resource Allocation Algorithm}
		\label{alg1}
		{	\begin{algorithmic}
				\STATE \textbf{Input:} $K, L, N, M, T_{f}, A_{\text{th}},  D_{\text{th}},\overline{{\cal P}},  \beta, \widehat{\alpha},\tau$, and $I$;
				\FOR {$i=1,\dots,  I$}
				\STATE Initialize the initial state $\bm{s}(1)$;
				\FOR {$\mu=1,\dots, N$}
				\IF {$rand(\cdot)<\tau$}
				\STATE Select a random action from $\textsf{A}$;
				\ELSE 
				\STATE Observe current state $\bm{s}(\mu)$ and select $\bm{a}(\mu)=\max\limits_{\bm{a}\in\textsf{A}}\left\{Q\left(\bm{s}(\mu),\bm{a}|\bm{\omega}_{\text{DDQN}}\right)\right\}$;
				\ENDIF
				\STATE BS broadcasts action $\bm{a}(\mu)$ and backlogged MDs execute the random access protocol;
				\STATE BS observes $\bm{s}(\mu+1)$ and computes   immediate reward; 
				\STATE Storing $\left(\bm{a}(\mu),\bm{s}(\mu),\bm{s}(\mu+1),R(\mu+1)\right)$ in $\Gamma$;
				\STATE Sampling  minibatch of transitions $\left(\bm{a}(j),\bm{s}(j),\bm{s}(j+1),R(j+1)\right)$ randomly from replay memory $\Gamma$;
				\STATE Calculate the loss of Q-function $\nabla\bm{{\cal L}}\left(\bm{\omega}_{\text{DDQN}}^{(\mu)}\right)$ by using Eq.~\eqref{equation200};
				\STATE Perform a gradient descent for each primary network and update the target networks using 
				$\overline{\bm{\omega}}_{\text{DDQN}}^{(\mu)}\leftarrow \psi\bm{\omega}_{\text{DDQN}}^{(\mu)}+(1-\psi)\overline{\bm{\omega}}_{\text{DDQN}}^{(\mu)}$.
				\ENDFOR
				\ENDFOR
		\end{algorithmic}}
	\end{algorithm}
	
		\begin{algorithm} [!tp]
		\algsetup{linenosize=\tiny} \small
		\caption{ DDNN-Based AoI-aware Resource Allocation Algorithm}
		\label{alg2}
		{	\begin{algorithmic}
				\STATE \textbf{Input:} $K, L, N, M, T_{f}, A_{\text{th}},  D_{\text{th}},\overline{{\cal P}}, \widehat{\alpha}, \beta, I,\xi,\chi$, and a threshold $\tau$;
				\STATE \textbf{Initialization:}  The primary network weights, target network weights, capacity $|\Gamma|$ and 
				the action-state value function $Q\left(\bm{s},\bm{a}|\overline{\bm{\omega}}_{\text{DDNN}},\xi,\chi\right)$
				\FOR {$i=1,\dots,  I$}
				\STATE Initialize the initial state $\bm{s}(1)$;
				\FOR {$\mu=1,\dots, N$}
				\IF {$rand(\cdot)<\tau$}
				\STATE Select a random action from $\textsf{A}$;
				\ELSE 
				\STATE Observe current state $\bm{s}(\mu)$ and select $\bm{a}(\mu)=\max\limits_{\bm{a}\in\textsf{A}}\left\{Q\left(\bm{s}(\mu),\bm{a}|\overline{\bm{\omega}}_{\text{DDNN}},\xi,\chi\right)\right\}$;
				\ENDIF
				\STATE BS broadcasts $\bm{a}(\mu)$, and backlogged MDs execute the random access protocol;
				\STATE BS observes $\bm{s}(\mu+1)$ and computes calculate the immediate reward; 
				\STATE Storing $\left(\bm{a}(\mu),\bm{s}(\mu),\bm{s}(\mu+1),R(\mu+1)\right)$ in $\Gamma$;
				\STATE Random sampling  minibatch of transitions $\left(\bm{a}(j),\bm{s}(j),\bm{s}(j+1),R(j+1)\right)$ from $\Gamma$;
				\STATE Combine the value function and advantage functions by using Eq.~\eqref{equation304};
				\STATE Compute the loss between the target and current Q-values 				
				\STATE Perform gradient descent to update the weights of the primary network based on the loss.
				\ENDFOR
				\ENDFOR
		\end{algorithmic}}
	\end{algorithm}
	
	\subsubsection{DDNN Algorithm}
	Due to the overestimation of optimizers, the convergence rate of the DDQN algorithm remains constrained, especially in large-scale systems~\cite{wang2016dueling}~\cite{8666109}. 
	To overcome such challenge, we apply the novel network slicing framework using the DDNN~\cite{wang2016dueling} to further improve the system's convergence speed. 
	Compared with conventional DRL approaches, the DDNN framework distinctly estimates the state values and the advantages of actions through two separate sequences of fully connected layers. This architecture enables the DDNN algorithm to obtain more robust estimates of state values, thereby significantly improving both its convergence rate and stability.
	We can derive the advantage function of actions, denoted by $G\left(\bm{s},\bm{a}(\mu)\right)$, as follows:
	\begin{equation}\label{equation202}
		G\left(\bm{s},\bm{a}(\mu)\right)=Q\left(\bm{s},\bm{a}(\mu)\right)-\widetilde{V}(\bm{s})
	\end{equation}
	where $\widetilde{V}(\bm{s})$ is specified by Eq.~\eqref{equation302}.
The Q-function $Q\left(\bm{s},\bm{a}(\mu)\right)$ indicates the value of action selection as in $\bm{s}$. 
The advantage function decouples the state value from the Q-function, thereby providing a relative measure of the significance of each action within a given state. To estimate the expected value function $\widetilde{V}(\bm{s})$ and the advantage function $G\left(\bm{s},\bm{a}(\mu)\right)$, we employ a dueling neural network architecture.
In this architecture, one sequence of fully connected layers is designed to output a scalar representation of the state value, denoted as  $\widetilde{V}(\bm{s}|\overline{\bm{\omega}}_{\text{DDNN}},\chi)$.
Concurrently, another sequence generates an 	$|\textsf{A}|$-dimensional vector that corresponds to the advantage function of actions, expressed as $G\left(\bm{s},\bm{a}(\mu)|\overline{\bm{\omega}}_{\text{DDNN}},\xi\right)$, where $\overline{\bm{\omega}}_{\text{DDNN}}$ represents the weight matrix of the convolutional layers and $\chi$ and $\xi$ are the parameters of fully-connected layers. 
	Then, we formulate the combining module of the network as follows:
	\begin{align}\label{equation304}
	&Q\left(\bm{s},\bm{a}(\mu)|\overline{\bm{\omega}}_{\text{DDNN}},\xi,\chi\right)=\widetilde{V}(\bm{s}|\overline{\bm{\omega}}_{\text{DDNN}},\chi)
		\nonumber\\
		&\qquad+\bigg[G\left(\bm{s},\bm{a}(\mu)|\overline{\bm{\omega}}_{\text{DDNN}},\xi\right)-\frac{1}{|\textsf{A}|}\sum_{\bm{a}\in\textsf{A}}G\left(\bm{s},\bm{a}|\overline{\bm{\omega}}_{\text{DDNN}},\xi\right)\!\bigg]\!.
	\end{align}
	As the dueling architecture shares the same input-output interface with DQN, we propose a DDNN-based AoI-aware resource allocation  algorithm, as illustrated in \textbf{Algorithm~\ref{alg2}}.
	This approach leverages the strengths of the dueling architecture to enhance the estimation of state values and action advantages, thereby improving the overall performance of the resource allocation in the context of mURLLC.

	\section{Performance Evaluations}\label{sec:results}

	We provide a set of numerical results to validate and evaluate our proposed DRL-based AoI-aware resource allocation schemes under multi-QoS through FBC. 
In our simulations, we model a scenario where MDs are randomly and uniformly distributed within a circular area with 500 m radius. Specifically, let the number of MDs be 
$K=50$, the status update packet size be $M =10^7$ bits, the duration of each transmission frame be $T_{f}=0.1$ s, the bandwidth of each subchannel be $B_{l}=100$ kHz, the number of subchannels be $L=4$, and the blocklength be $n_{l}\in[100,600]$. 
	
	\begin{figure}[!t]
		\centering
		\includegraphics[scale=0.4]{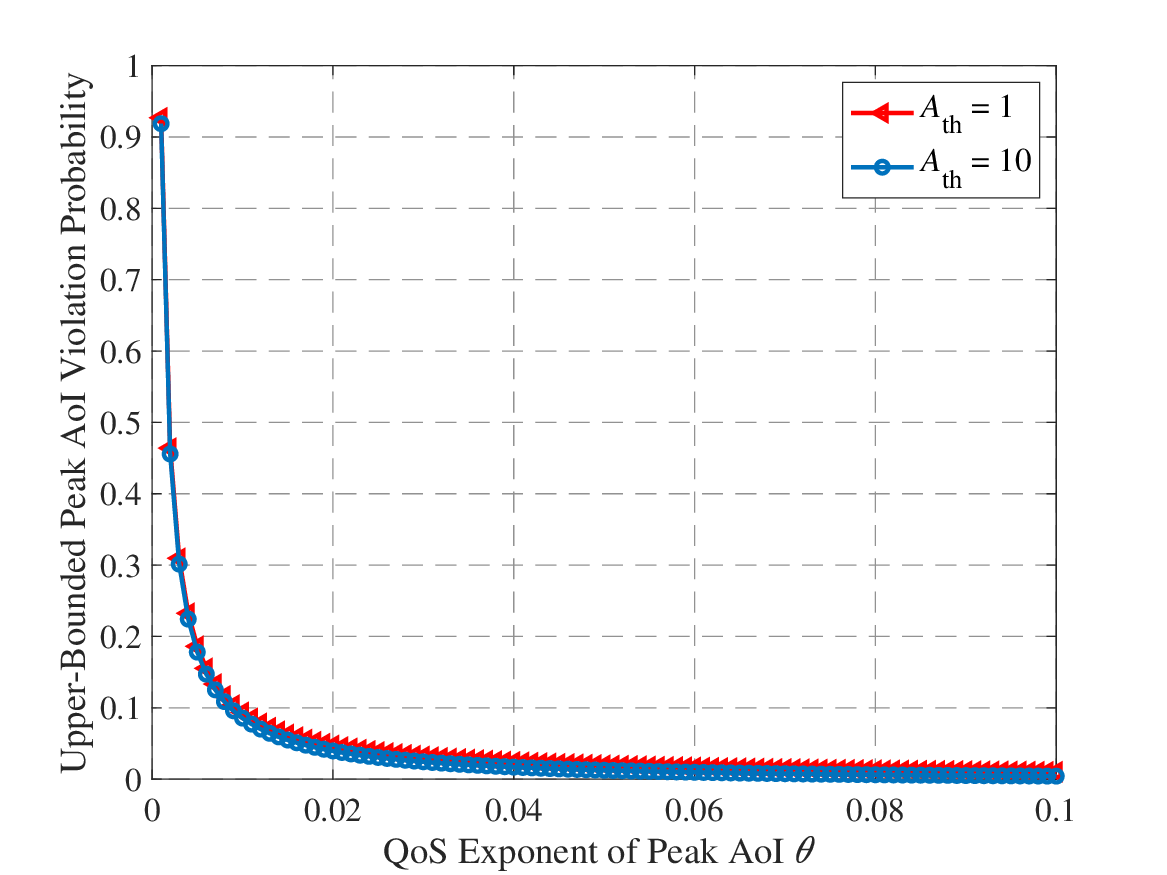}
		\caption{The upper-bounded peak AoI violation probability against QoS exponent of peak AoI $\theta$.}
		\label{fig:03}
	\end{figure}
	
Now we set the blocklength $n_{l}=400$ and average SNR to be $-5$ dB. Fig.~\ref{fig:03} illustrates the relationship between the upper-bound of the peak AoI violation probability and the QoS exponent of peak AoI $\theta$ for AoI-aware schemes.
As illustrated in Fig.~\ref{fig:03}, an increase in the peak AoI threshold results in a reduction in the upper-bounded peak AoI violation probability. Specifically, Fig.~\ref{fig:03} demonstrates that the upper-bounded peak AoI violation probability decreases with an increase in $\theta$.
This behavior indicates that 
a smaller QoS exponent for peak AoI imposes an upper bound, whereas a larger QoS exponent establishes a lower bound on the upper-bounded peak AoI violation probability.

	\begin{figure}[!t]
		\centering
		\includegraphics[scale=0.4]{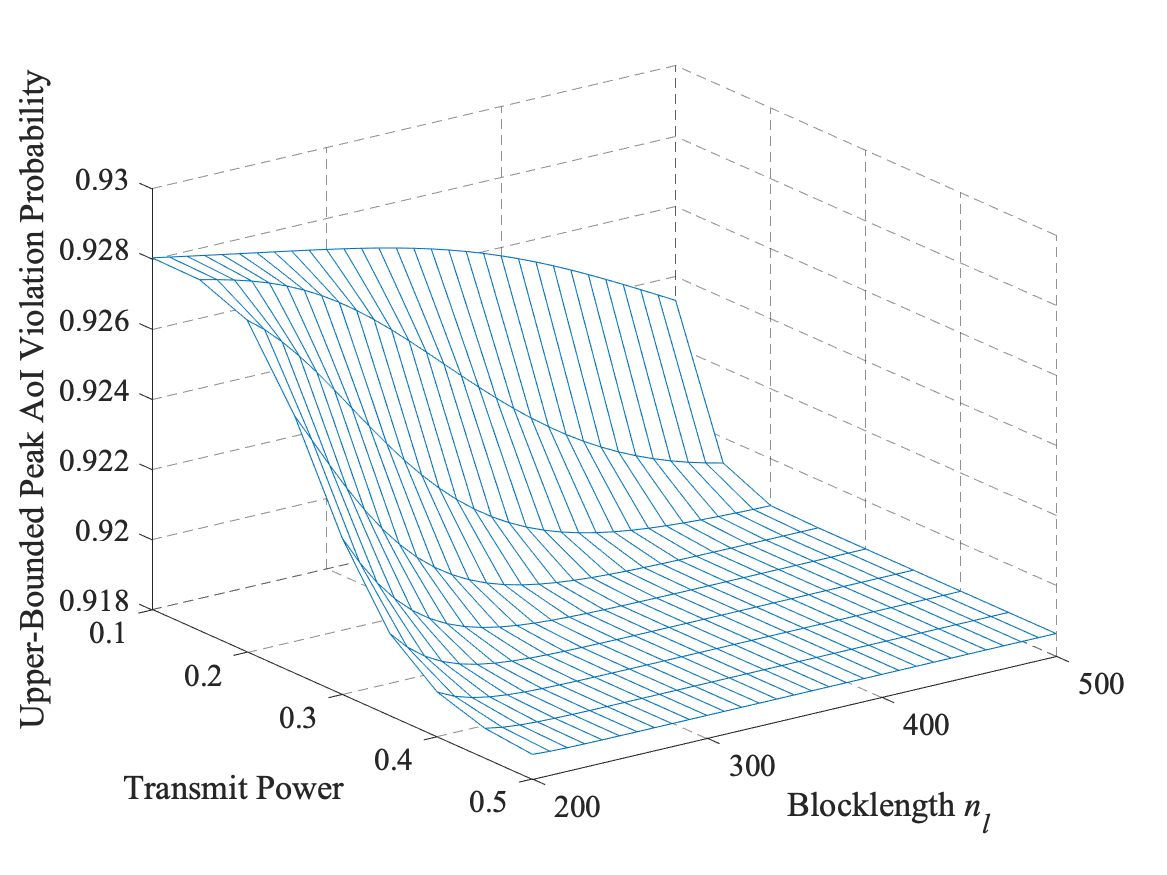}
		\caption{The upper-bounded peak AoI violation probability against blocklength and transmit power through FBC.}
		\label{fig:04}
	\end{figure}
	
	Now we set the peak AoI threshold $A_{\text{th}}=10$ and the QoS exponent of peak AoI $\theta=0.001$.   Fig.~\ref{fig:04} plots the upper-bounded peak AoI violation probability $p^{(\mu,\text{AoI})}_{k,l}$ as a function of both the blocklength and the transmit power using FBC.
	We can observe from Fig.~\ref{fig:04} that the upper-bounded peak AoI violation probability is a monotonically decreasing function with respect to blocklength and transmit power, implying that we need to allocate more power and blocklength for achieving a better performance in terms of the peak AoI violation probability.

	\begin{figure}[!t]
		\centering
		\includegraphics[scale=0.4]{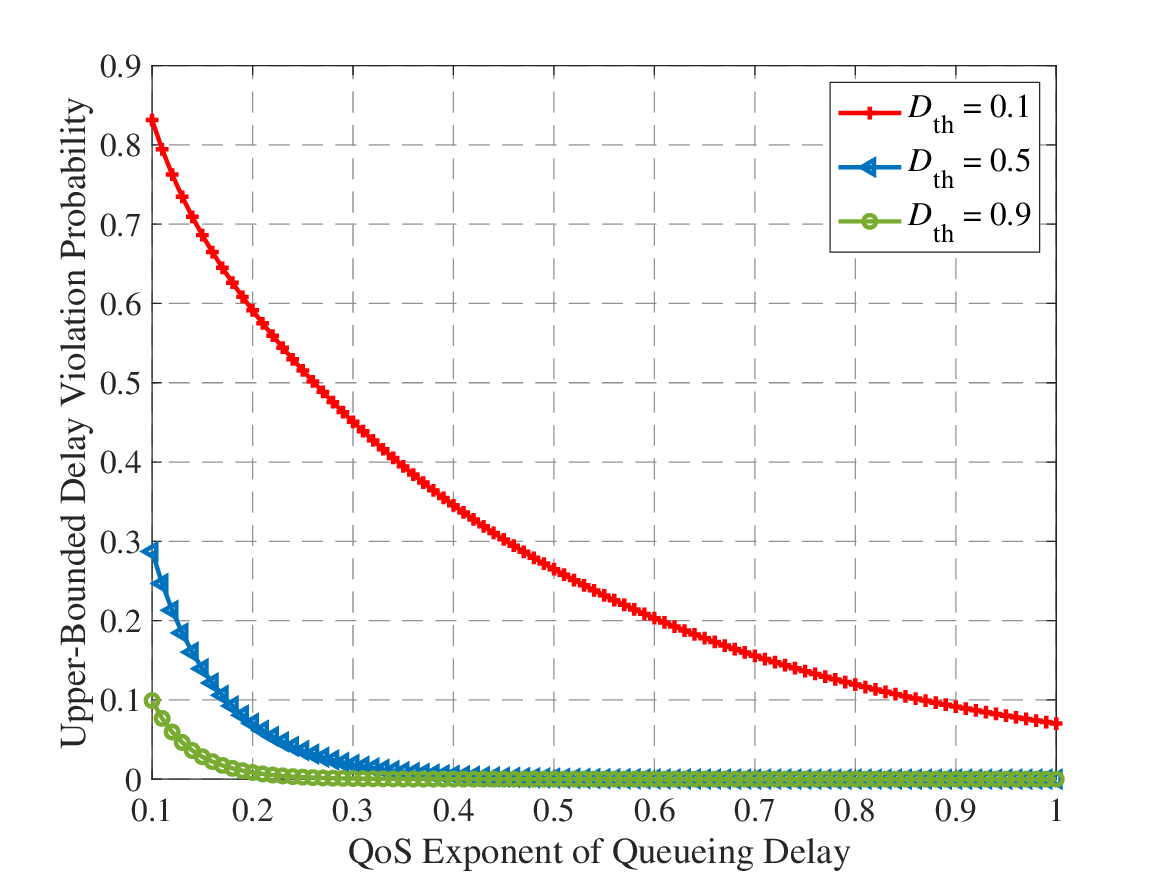}
		\caption{The upper-bounded delay violation probability $p^{(\mu,\text{q})}_{k,l}$ against QoS exponent of queueing delay $\widetilde{\theta}$.}
		\label{fig:05}
	\end{figure}
	
	Setting the average SNR to be $1$ dB and blocklength $n_{l}=500$,  Fig.~\ref{fig:05} plots the upper-bounded delay violation probability $p^{(\mu,\text{q})}_{k,l}$ as a function of the QoS exponent of queueing delay $\widetilde{\theta}$.
As depicted in Fig.~\ref{fig:05}, a larger delay bound leads to a reduction in the upper-bounded delay violation probability. 
This illustrates upper-bound on the delay violation probability decreases when the QoS exponent of queueing delay $\widetilde{\theta}$ increases. This indicates that when $\widetilde{\theta} \rightarrow 0$, the proposed scheme is capable of tolerating longer delays. 
Conversely, as $\widetilde{\theta} \rightarrow \infty$, the system's tolerance for delay becomes more constrained, allowing only very short delays.	

	\begin{figure}[!t]
		\centering
		\includegraphics[scale=0.4]{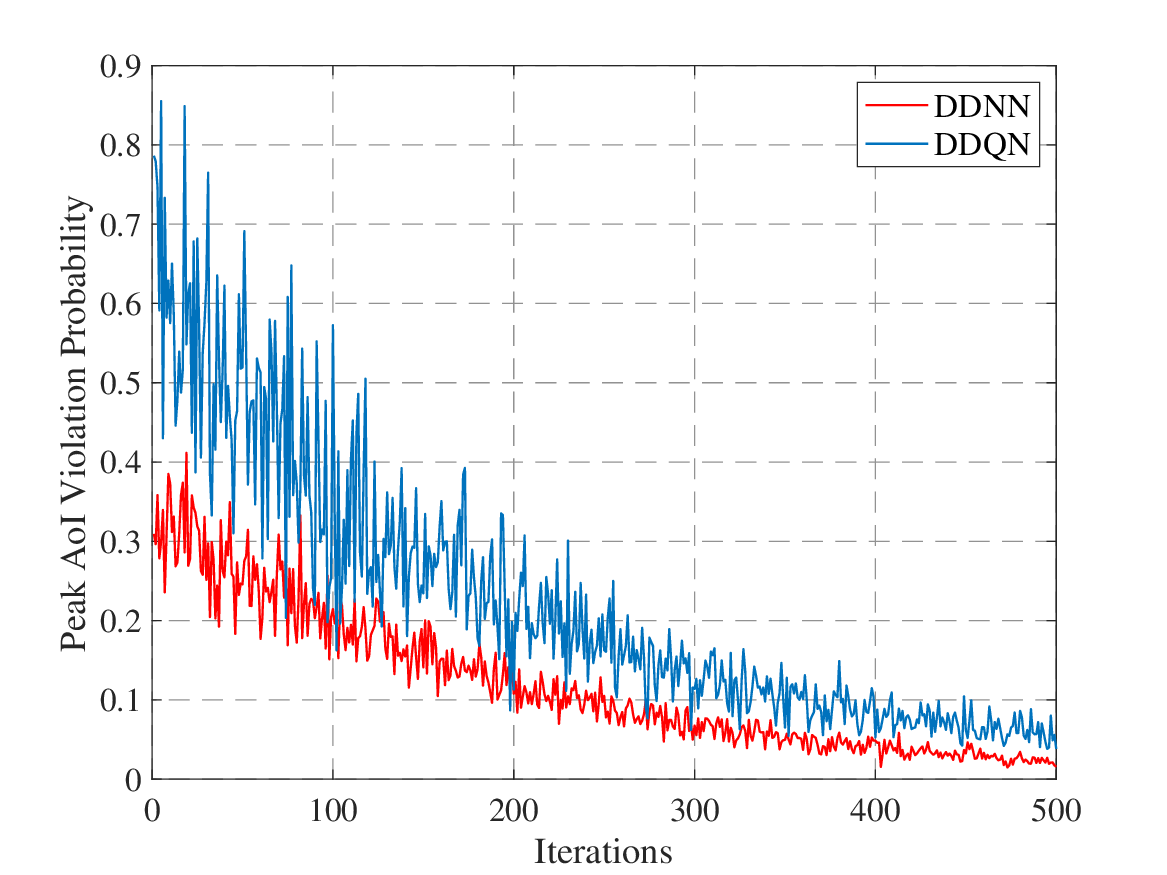}
		\caption{The peak AoI violation probability $p^{(\mu,\text{AoI})}_{k,l}$ against the number of iterations with $\theta=0.01$.}
		\label{fig:06}
	\end{figure}
	
		\begin{figure}[!t]
		\centering
		\includegraphics[scale=0.4]{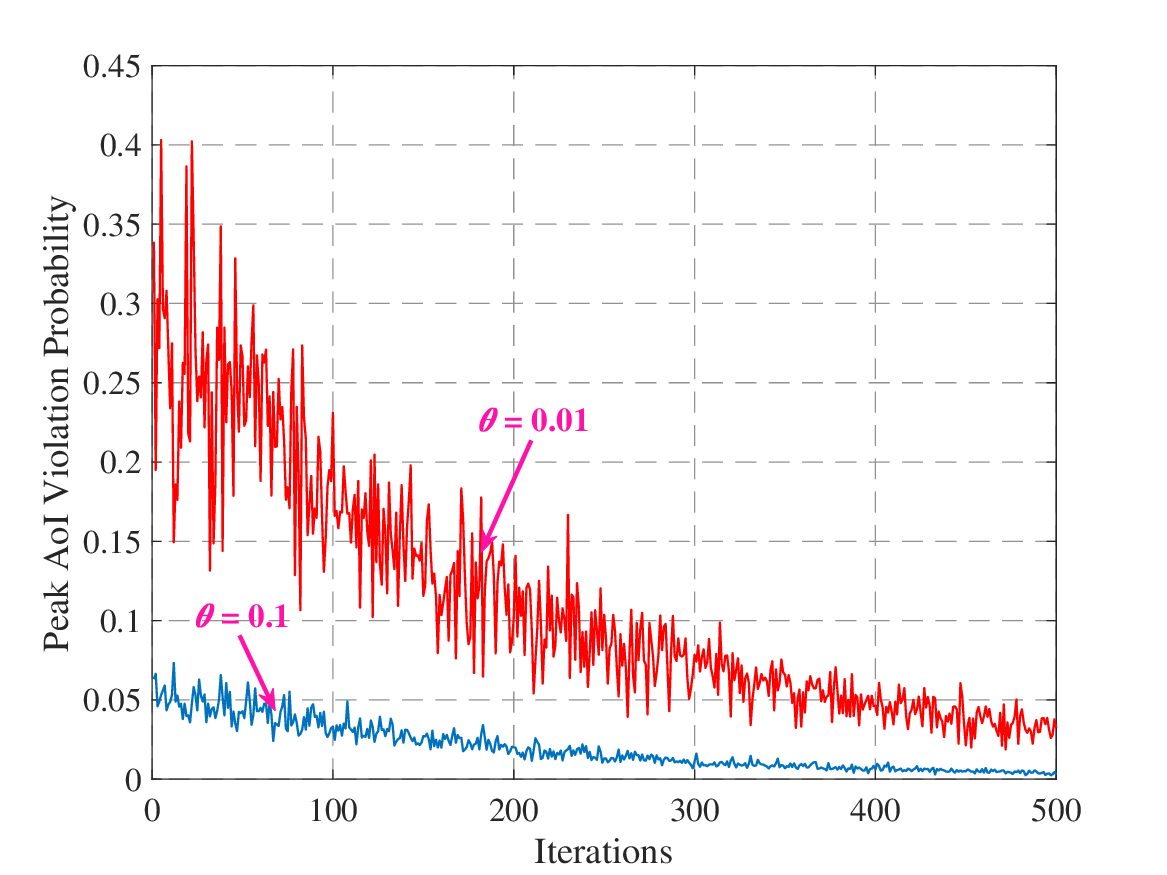}
		\caption{The peak AoI violation probability $p^{(\mu,\text{AoI})}_{k,l}$ against the number of iterations with $\theta=\{0.01,0.1\}$.}
		\label{fig:06b}
	\end{figure}

	To compare the peak AoI performance for our proposed DRL-based algorithms, Fig.~\ref{fig:06} presents the implementation of DDQN and DDNN algorithms proposed in \textbf{Algorithm~\ref{alg1}} and \textbf{Algorithm~\ref{alg2}}, respectively, for analyzing the peak AoI violation probability against the executed sequence of training iterations with the QoS exponent of peak AoI $\theta=0.01$.
	We can also observe from Fig.~\ref{fig:06} that the DDNN-based AoI-aware resource allocation algorithm outperforms the DDQN algorithm in terms of the peak AoI violation probability. 
	In addition, Fig.~\ref{fig:06b} plots the peak AoI violation probability  $p^{(\mu,\text{AoI})}_{k,l}$ DDNN algorithm proposed in \textbf{Algorithm~\ref{alg2}} after experiencing sequence of training iterations with different QoS exponents of peak AoI $\theta=0.01$ and $\theta=0.1$, respectively. 
	From Fig.~\ref{fig:06b} we observe that a larger QoS exponent of peak AoI $\theta$ can lead to a lower peak AoI violation probability. 
	
	\begin{figure}[!t]
	\centering
	\includegraphics[scale=0.404]{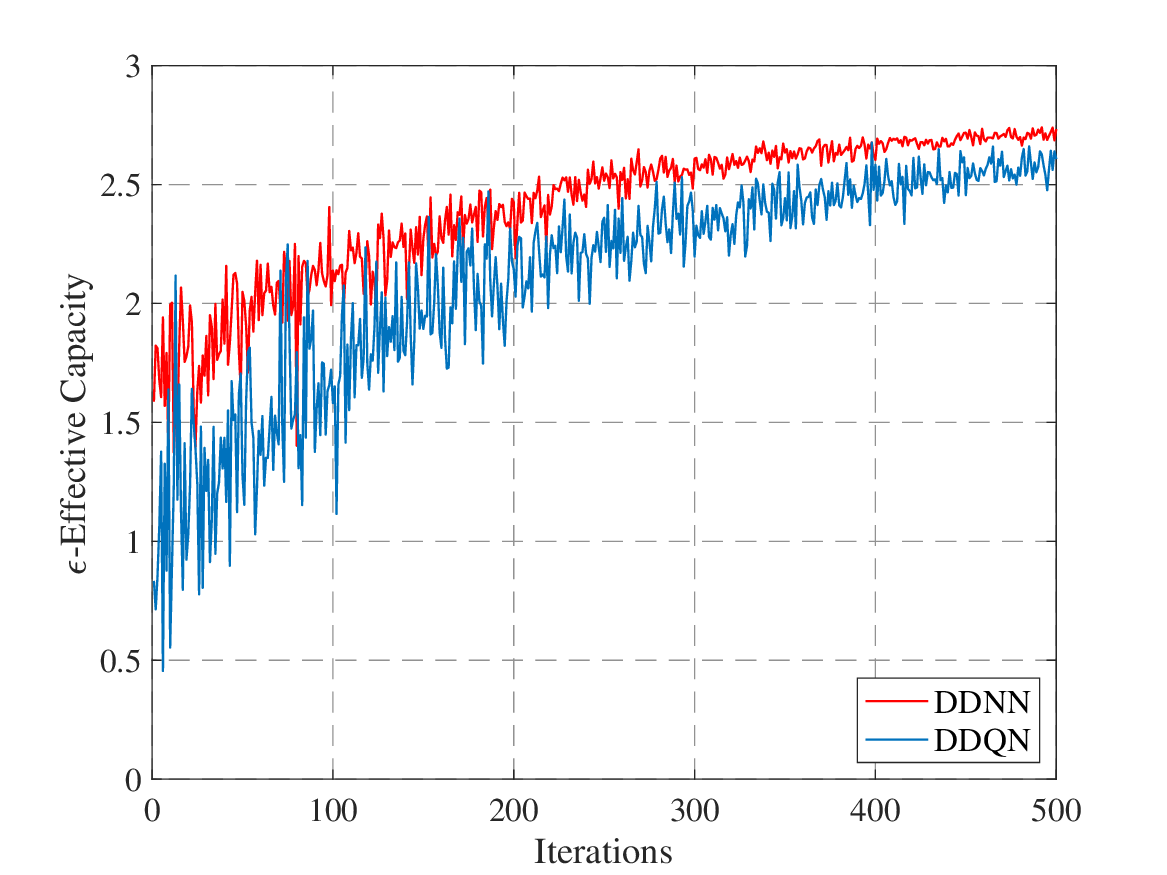}
	\caption{The $\epsilon$-effective capacity $EC_{k,l}^{(\epsilon,\mu)}$ against the number of iterations through FBC}
	\label{fig:07}
\end{figure}
	
	In addition, setting the QoS exponent of queuing delay $\widetilde{\theta}=0.1$, Fig.~\ref{fig:07} presents the implementation of DDQN and DDNN algorithms for analyzing the $\epsilon$-effective capacity $EC_{k,l}^{(\epsilon,\mu)}$ as the function of the sequence of training iterations. 
Fig.~\ref{fig:07} illustrates that DDNN-based AoI-aware resource allocation algorithm yields the better perform than the DDQN-based AoI-aware resource allocation algorithm in terms of the $\epsilon$-effective capacity.

	\section{Conclusions}\label{sec:conclusion} 
	We have proposed the DRL-based AoI-aware resource allocations for multi-QoS provisioning schemes through FBC. 
	In particular, we have developed network architecture communications models, AoI-metric based modeling frameworks, and random access protocol to support mURLLC using FBC.
	We also have derived the closed-form expressions of upper-bounds for the AoI violation probability and delay violation probability functions by applying the SNC in supporting our proposed statistical delay and error-rate bounded QoS provisioning in the finite blocklength regime. 
	We have formulated the AoI-aware resource allocation optimization problem, modeled the optimization problem by MDP, and applied two DRL techniques, including DDQN and DDNN algorithms in supporting multi-QoS for mURLLC.
	A series of simulations have been conducted to verify, assess, and examine our established multi-QoS provisioning schemes.

	\nocite{*}
	\footnotesize
	\bibliographystyle{IEEEtran}
	\bibliography{myref.bib}

\begin{thebibliography}{10}
\providecommand{\url}[1]{#1}
\csname url@samestyle\endcsname
\providecommand{\newblock}{\relax}
\providecommand{\bibinfo}[2]{#2}
\providecommand{\BIBentrySTDinterwordspacing}{\spaceskip=0pt\relax}
\providecommand{\BIBentryALTinterwordstretchfactor}{4}
\providecommand{\BIBentryALTinterwordspacing}{\spaceskip=\fontdimen2\font plus
\BIBentryALTinterwordstretchfactor\fontdimen3\font minus
  \fontdimen4\font\relax}
\providecommand{\BIBforeignlanguage}[2]{{%
\expandafter\ifx\csname l@#1\endcsname\relax
\typeout{** WARNING: IEEEtran.bst: No hyphenation pattern has been}%
\typeout{** loaded for the language `#1'. Using the pattern for}%
\typeout{** the default language instead.}%
\else
\language=\csname l@#1\endcsname
\fi
#2}}
\providecommand{\BIBdecl}{\relax}
\BIBdecl

\bibitem{10494937}
J.~Wang, W.~Cheng, W.~Zhang, and H.~Liang, ``Statistical {QoS} provisioning
  architecture for {6G} satellite-terrestrial integrated networks,''
  \emph{Journal of Communications and Information Networks}, vol.~9, no.~1, pp.
  34--42, 2024.

\bibitem{10054381}
C.-X. Wang, X.~You, X.~Gao, X.~Zhu, Z.~Li, C.~Zhang, H.~Wang, Y.~Huang,
  Y.~Chen, H.~Haas, J.~S. Thompson, E.~G. Larsson, M.~D. Renzo, W.~Tong,
  P.~Zhu, X.~Shen, H.~V. Poor, and L.~Hanzo, ``On the road to {6G}: Visions,
  requirements, key technologies, and testbeds,'' \emph{IEEE Communications
  Surveys \& Tutorials}, vol.~25, no.~2, pp. 905--974, 2023.

\bibitem{10177877}
Y.~Zhang, W.~Cheng, and W.~Zhang, ``Multiple access integrated adaptive finite
  blocklength for ultra-low delay in {6G} wireless networks,'' \emph{IEEE
  Transactions on Wireless Communications}, vol.~23, no.~3, pp. 1670--1683,
  2024.

\bibitem{9635675}
W.~Cheng, Y.~Xiao, S.~Zhang, and J.~Wang, ``Adaptive finite blocklength for
  ultra-low latency in wireless communications,'' \emph{IEEE Transactions on
  Wireless Communications}, vol.~21, no.~6, pp. 4450--4463, 2022.

\bibitem{7529226}
G.~{Durisi}, T.~{Koch}, and P.~{Popovski}, ``Toward massive, ultrareliable, and
  low-latency wireless communication with short packets,'' \emph{Proceedings of
  the IEEE}, vol. 104, no.~9, pp. 1711--1726, 2016.

\bibitem{yury2010}
Y.~Polyanskiy, H.~V. Poor, and S.~Verd\'{u}, ``Channel coding rate in the
  finite blocklength regime,'' \emph{IEEE Transactions on Information Theory},
  vol.~56, no.~5, pp. 2307--2359, May 2010.

\bibitem{Yp2014}
Y.~Polyanskiy and S.~Verd{\'u}, ``Empirical distribution of good channel codes
  with non-vanishing error probability,'' \emph{IEEE transactions on
  information theory}, vol.~60, no.~1, pp. 5--21, 2013.

\bibitem{chen2020massive}
X.~Chen, D.~W.~K. Ng, W.~Yu, E.~G. Larsson, N.~Al-Dhahir, and R.~Schober,
  ``Massive access for {5G} and beyond,'' \emph{IEEE Journal on Selected Areas
  in Communications}, vol.~39, no.~3, pp. 615--637, 2021.

\bibitem{10334482}
X.~Zhang, Z.~Chang, T.~H{\"a}m{\"a}l{\"a}inen, and G.~Min, ``{AoI}-energy
  tradeoff for data collection in {UAV}-assisted wireless networks,''
  \emph{IEEE Transactions on Communications}, vol.~72, no.~3, pp. 1849--1861,
  2024.

\bibitem{9023932}
Z.~Jiang, S.~Fu, S.~Zhou, Z.~Niu, S.~Zhang, and S.~Xu, ``{AI}-assisted low
  information latency wireless networking,'' \emph{IEEE Wireless
  Communications}, vol.~27, no.~1, pp. 108--115, 2020.

\bibitem{8691802}
J.-B. Seo and J.~Choi, ``On the outage probability of peak age-of-information
  for {D/G/1} queuing systems,'' \emph{IEEE Communications Letters}, vol.~23,
  no.~6, pp. 1021--1024, 2019.

\bibitem{9145084}
B.~Han, Z.~Jiang, Y.~Zhu, and H.~D. Schotten, ``Recursive optimization of
  finite blocklength allocation to mitigate age-of-information outage,'' in
  \emph{2020 IEEE International Conference on Communications Workshops (ICC
  Workshops)}, 2020, pp. 1--6.

\bibitem{abdel2018ultra}
M.~K. Abdel-Aziz, C.-F. Liu, S.~Samarakoon, M.~Bennis, and W.~Saad,
  ``Ultra-reliable low-latency vehicular networks: Taming the age of
  information tail,'' in \emph{2018 IEEE Global Communications Conference
  (GLOBECOM)}.\hskip 1em plus 0.5em minus 0.4em\relax IEEE, 2018, pp. 1--7.

\bibitem{8437671}
R.~Devassy, G.~Durisi, G.~C. Ferrante, O.~Simeone, and E.~Uysal-Biyikoglu,
  ``Delay and peak-age violation probability in short-packet transmissions,''
  in \emph{2018 IEEE International Symposium on Information Theory (ISIT)},
  2018, pp. 2471--2475.

\bibitem{8402240}
Y.~Hu, M.~Ozmen, M.~C. Gursoy, and A.~Schmeink, ``Optimal power allocation for
  {QoS}-constrained downlink multi-user networks in the finite blocklength
  regime,'' \emph{IEEE Transactions on Wireless Communications}, vol.~17,
  no.~9, pp. 5827--5840, 2018.

\bibitem{sutton2018reinforcement}
R.~S. Sutton and A.~G. Barto, \emph{Reinforcement Learning: An
  Introduction}.\hskip 1em plus 0.5em minus 0.4em\relax MIT press, 2018.

\bibitem{8664581}
N.~Jiang, Y.~Deng, A.~Nallanathan, and J.~A. Chambers, ``Reinforcement learning
  for real-time optimization in {NB-IoT} networks,'' \emph{IEEE Journal on
  Selected Areas in Communications}, vol.~37, no.~6, pp. 1424--1440, 2019.

\bibitem{8743390}
Y.~Sun, M.~Peng, Y.~Zhou, Y.~Huang, and S.~Mao, ``Application of machine
  learning in wireless networks: Key techniques and open issues,'' \emph{IEEE
  Communications Surveys \& Tutorials}, vol.~21, no.~4, pp. 3072--3108, 2019.

\bibitem{10145068}
A.~Zakeri, M.~Moltafet, M.~Leinonen, and M.~Codreanu, ``Minimizing the {AoI} in
  resource-constrained multi-source relaying systems: Dynamic and
  learning-based scheduling,'' \emph{IEEE Transactions on Wireless
  Communications}, vol.~23, no.~1, pp. 450--466, 2024.

\bibitem{6226795}
F.~Capozzi, G.~Piro, L.~Grieco, G.~Boggia, and P.~Camarda, ``Downlink packet
  scheduling in {LTE} cellular networks: Key design issues and a survey,''
  \emph{IEEE Communications Surveys \& Tutorials}, vol.~15, no.~2, pp.
  678--700, 2013.

\bibitem{8006544}
R.~D. Yates and S.~K. Kaul, ``Status updates over unreliable multiaccess
  channels,'' in \emph{2017 IEEE International Symposium on Information Theory
  (ISIT)}, 2017, pp. 331--335.

\bibitem{8445979}
R.~Talak, S.~Karaman, and E.~Modiano, ``Distributed scheduling algorithms for
  optimizing information freshness in wireless networks,'' in \emph{2018 IEEE
  19th International Workshop on Signal Processing Advances in Wireless
  Communications (SPAWC)}, 2018, pp. 1--5.

\bibitem{9181539}
B.~Yu, Y.~Cai, and D.~Wu, ``Joint access control and resource allocation for
  short-packet-based {mMTC} in status update systems,'' \emph{IEEE Journal on
  Selected Areas in Communications}, vol.~39, no.~3, pp. 851--865, 2021.

\bibitem{7404058}
S.~Duan, V.~Shah-Mansouri, Z.~Wang, and V.~W.~S. Wong, ``{D-ACB}: Adaptive
  congestion control algorithm for bursty m2m traffic in lte networks,''
  \emph{IEEE Transactions on Vehicular Technology}, vol.~65, no.~12, 2016.

\bibitem{6398917}
H.~Wu, C.~Zhu, R.~J. La, X.~Liu, and Y.~Zhang, ``Fast adaptive {S-ALOHA} scheme
  for event-driven machine-to-machine communications,'' in \emph{2012 IEEE
  Vehicular Technology Conference (VTC Fall)}, 2012.

\bibitem{10609803}
J.~Wang, W.~Cheng, and H.~Vincent~Poor, ``Statistical delay and error-rate
  bounded {QoS} provisioning for {AoI}-driven {6G} satellite-terrestrial
  integrated networks using {FBC},'' \emph{IEEE Transactions on Wireless
  Communications}, 2024.

\bibitem{HAL2016}
H.~{Al-Zubaidy}, J.~{Liebeherr}, and A.~{Burchard}, ``Network-layer performance
  analysis of multihop fading channels,'' \emph{IEEE/ACM Transactions on
  Networking}, vol.~24, no.~1, pp. 204--217, Feb. 2016.

\bibitem{JSAC_jingqing2021}
X.~Zhang, J.~Wang, and H.~V. Poor, ``{AoI}-driven statistical delay and
  error-rate bounded {QoS} provisioning for {6G mURLLC} over {UAV} multimedia
  mobile networks using {FBC},'' \emph{IEEE Journal on Selected Areas in
  Communications}, Accepted to appear in July 2021.

\bibitem{9400231}
------, ``{AoI}-driven statistical delay and error-rate bounded {QoS}
  provisioning for {URLLC} in the finite blocklength regime,'' in \emph{2021
  55th Annual Conference on Information Sciences and Systems (CISS)}, 2021, pp.
  1--6.

\bibitem{10097504}
M.~Dampfhoffer, T.~Mesquida, A.~Valentian, and L.~Anghel,
  ``Backpropagation-based learning techniques for deep spiking neural networks:
  A survey,'' \emph{IEEE Transactions on Neural Networks and Learning Systems},
  vol.~35, no.~9, pp. 11\,906--11\,921, 2024.

\bibitem{9763399}
C.~Lee, J.~Jung, and J.-M. Chung, ``Intelligent dual active protocol stack
  handover based on double {DQN} deep reinforcement learning for {5G} {mmWave}
  networks,'' \emph{IEEE Transactions on Vehicular Technology}, vol.~71, no.~7,
  pp. 7572--7584, 2022.

\bibitem{LI2024127204}
X.~Li, B.~Tang, and H.~Li, ``{AdaER}: An adaptive experience replay approach
  for continual lifelong learning,'' \emph{Neurocomputing}, vol. 572, p.
  127204, 2024.

\bibitem{wang2016dueling}
Z.~Wang, T.~Schaul, M.~Hessel, H.~Hasselt, M.~Lanctot, and N.~Freitas,
  ``Dueling network architectures for deep reinforcement learning,'' in
  \emph{International conference on machine learning}.\hskip 1em plus 0.5em
  minus 0.4em\relax PMLR, 2016, pp. 1995--2003.

\bibitem{8666109}
N.~Van~Huynh, D.~Thai~Hoang, D.~N. Nguyen, and E.~Dutkiewicz, ``Optimal and
  fast real-time resource slicing with deep dueling neural networks,''
  \emph{IEEE Journal on Selected Areas in Communications}, vol.~37, no.~6, pp.
  1455--1470, 2019.

\end{thebibliography}

\end{document}